%% file: noise-rec.tex
\documentclass[pre,reprint,amsmath,showpacs,showkeys,nofootinbib]{revtex4-1}

\usepackage{graphicx,amsfonts}
\usepackage{bbold}
\usepackage{hyperref,ifthen}

\begin{document}

\title{Reconstructing signals from noisy data with unknown signal and noise covariance}
\author{Niels Oppermann}\email{niels@mpa-garching.mpg.de}
\author{Georg Robbers}
\author{Torsten A. En{\ss}lin}
\affiliation{Max Planck Institute for Astrophysics, Karl-Schwarzschild-Str. 1, 85741 Garching, Germany}
\date{\today}


\begin{abstract}
	We derive a method to reconstruct Gaussian signals from linear measurements with Gaussian noise. This new algorithm is intended for applications in astrophysics and other sciences. The starting point of our considerations is the principle of minimum Gibbs free energy which was previously used to derive a signal reconstruction algorithm handling uncertainties in the signal covariance. We extend this algorithm to simultaneously uncertain noise and signal covariances using the same principles in the derivation. The resulting equations are general enough to be applied in many different contexts. We demonstrate the performance of the algorithm by applying it to specific example situations and compare it to algorithms not allowing for uncertainties in the noise covariance. The results show that the method we suggest performs very well under a variety of circumstances and is indeed qualitatively superior to the other methods in cases where uncertainty in the noise covariance is present.
\end{abstract}

\maketitle

\section{Introduction}

The problem of signal inference consists of reconstructing a set of parameters or even a continuous field $s$ from some data set $d$, which is influenced in some way by the signal,
\begin{equation}
	\label{eq:datamodel-general}
	d=f(s)+n.
\end{equation}
Two problems will arise. First, the function $f$ may not be invertible and, second, the noise term $n$ will not be known. In the Bayesian framework, one uses prior information on the signal and the noise term to calculate a best estimate for the true signal realization or, ideally, the whole probability distribution for the signal given the prior information and the information contained in the data.

Symmetry considerations and knowledge about the underlying physics of the signal and the measurement process may restrict the class of priors that one has to consider. They might, however, still contain some free parameters that then become part of the inference problem. The case in which the signal covariance contains uncertain parameters was tackled in \cite{ensslin_frommert-2011}, producing a whole class of filters for this problem. The filter that we extend in this work was reproduced in \cite{ensslin_weig-2010}, where the principle of minimum Gibbs free energy was introduced (cf. also Sect.~\ref{sec:derivation}), and successfully applied in an astrophysical setting in \cite{oppermann-2011a}.

Here, we focus on the case where we can assume zero-mean Gaussian priors both for the signal and for the noise. The priors are therefore completely characterized by the respective covariance matrices. Our goal is to extend the study of \cite{ensslin_weig-2010} to the case in which both the signal covariance and the noise covariance contain parameters that are not known a priori. This is motivated mainly by applications from the field of astrophysics. The theory and resulting filter formulas, however, are of general applicability. Gaussian noise, e.g., is omnipresent in nearly every area of the natural sciences and the situation in which its variance is not precisely known should be a rather common one.

Previous work dealing with the problem of unknown noise variance has mainly dealt with specific applications. One of these applications is the field of image reconstruction. Here, it is usually assumed that the measured picture is the sum of the underlying signal and a white Gaussian noise term. Often, it is further assumed that the noise level, i.e. its variance, is the same in every image pixel. A comparison of different algorithms for noise estimation under these assumptions can be found e.g. in \cite{olsen-1993}. An example for an algorithm allowing for inhomogeneous noise is presented in \cite{starck-1998}, where a wavelet transform of the image is applied and the lack of correlated noise is exploited. Most of these algorithms, however, are not derived by rigorous statistical calculations but rather by a combination of intuition and experience.

From a mathematical viewpoint, the problem of an unknown noise prior has received some attention in the theory of density deconvolution, which deals with the inference of the probability density for a signal from measurements with additive noise. Here, the signal is usually assumed to consist of independent identically distributed variables. The case of Gaussian noise with unknown variance has been considered e.g. by \cite{koltchinskii-2000} and \cite{schwarz-2010}.

In this work, we create a general setting with well defined assumptions and a traceable derivation of a general filter formula within a Bayesian framework, not loosing sight of its applicability. Our result can accomodate a host of different assumptions and models, such as correlated or uncorrelated noise. It allows for a distinction between the data space and the signal space, with possibly different numbers of degrees of freedom. 

The remainder of the paper is organized as follows. In Sect.~\ref{sec:notation} we introduce our model for the measurement process and the notation that is required. The derivation of the filter formulas follows in Sect.~\ref{sec:derivation}. We then demonstrate the usefulness of our filter by applying it in a set of mock observational situations in Sect.~\ref{sec:application} and discuss the implications in Sect.~\ref{sec:discussion}.

\section{Signal model and notation}
\label{sec:notation}

We assume a linear measurement process where the data are a superposition of a linear signal response and a noise term,
\begin{equation}
	d=Rs+n.
\end{equation}
Here, both the data and noise and the signal can be finite-dimensional vectors or continuous fields defined on some manifold. The response matrix $R$ maps a field in the signal space to a field in the data space. In the continuous limit the matrix vector product becomes
\begin{equation}
	\left(Rs\right)_i=\int\mathrm{d}x~R_{ix}s_x,
\end{equation}
where the index denotes the value of a field at this position. In physical applications the data vector will always be discrete, since only a finite number of measurements can be taken, while the signal space might well be continuous. The result of any numerical signal reconstruction, however, will at best be a discretized version of the continuous field.

Further, we assume Gaussian prior statistics both for the signal and for the noise contribution, i.e. $s\hookleftarrow\mathcal{G}(s,S)$, $n\hookleftarrow\mathcal{G}(n,N)$, where
\begin{equation}
	\mathcal{G}(a,A)=\frac{1}{\sqrt{\left|2\pi A\right|}}\exp\left(-\frac{1}{2}a^\dagger A^{-1}a\right)
\end{equation}
denotes a multivariate Gaussian distribution in $a$ with covariance matrix $A$. We use the dagger symbol to indicate a scalar product,
\begin{equation}
	a^\dagger b=\int\mathrm{d}x~a_x^*b_x,
\end{equation}
and the asterisk to denote complex conjugation. This corresponds to the notation introduced in \cite{ensslin_frommert_kitaura-2009}.

The problem of signal reconstruction is to find an optimal estimate $m$ for the signal realization that the measured data arose from. Optimality in an $L^2$-norm sense leads to
\begin{equation}
	m=\left<s\right>_{\mathcal{P}(s|d)}:=\int\mathcal{D}s~s\mathcal{P}(s|d),
\end{equation}
i.e. the posterior mean. The integration is performed over all possible signal configurations. In the discrete case this becomes a product of one-dimensional integrals,
\begin{equation}
	\int\mathcal{D}s=\int_{-\infty}^{+\infty}\mathrm{d}s_1\int_{-\infty}^{\infty}\mathrm{d}s_2\cdots,
\end{equation}
where $s=(s_1,s_2,\dots)$ is the vector of signal values at locations $1,2,\dots$. Ideally, we would also like to obtain some information on the posterior distribution $\mathcal{P}(s|d)$ other than its mean. If the signal and noise covariances are known, the posterior is a Gaussian $\mathcal{G}(s-m,D)$ with mean
\begin{equation}
	\label{eq:WF}
	m=Dj,
\end{equation}
and covariance $D$, where $j=R^\dagger N^{-1}d$ is called the \textit{information source} and $D=\left(S^{-1}+R^\dagger N^{-1}R\right)^{-1}$ the \textit{information propagator} \citep[cf.][]{ensslin_frommert_kitaura-2009} and the dagger attached to a matrix denotes its hermitian conjugate.

In this paper we are concerned with the case in which neither the signal covaricance matrix $S$ nor the noise covariance matrix $N$ are known. We parameterize these matrices as sums of their eigenvalues $\tilde{p}_k$ and $\tilde{\eta}_j$ multiplied with the projectors onto the respective eigenspaces $\tilde{S}_k$ and $\tilde{N}_j$. The parameters can be rescaled by including some numerical values $\tilde{s}_k$ and $\tilde{n}_j$ in the projection-like matrices, making the rescaled version of the parameterization
\begin{align}
	S&=\sum_kp_kS_k,\label{eq:Sdecomp}\\
	N&=\sum_j\eta_jN_j\label{eq:Ndecomp},
\end{align}
where
\begin{equation}
	p_k=\frac{\tilde{p}_k}{\tilde{s}_k},~\eta_j=\frac{\tilde{\eta}_j}{\tilde{n}_j}
\end{equation}
and
\begin{equation}
	S_k=\tilde{s}_k\tilde{S}_k,~N_j=\tilde{n}_j\tilde{N}_j.
\end{equation}
Furthermore, we define the pseudo-inverse matrices $S_k^{-1}=\tilde{s}_k^{-1}\tilde{S}_k$ and $N_j^{-1}=\tilde{n}_j^{-1}\tilde{N}_j$, so that $S_k^{-1}S_k$ and $N_j^{-1}N_j$ are identity operators on the respective eigenspaces.

We assume here that the eigenspaces corresponding to the different eigenvalues are known a priori, e.g. from symmetry considerations. However, the formalism allows for eigenvalues of different eigenspaces becoming equal a posteriori.

Finally, we also need to define some priors for the parameters $p_k$ and $\eta_j$. As was done in \cite{ensslin_frommert-2011} and \cite{ensslin_weig-2010}, we assume each parameter to be a priori independent from all the others and use inverse Gamma distributions, i.e. power laws with exponential cutoff, as priors for the individual parameters,
\begin{eqnarray}
	\mathcal{P}(p,\eta)&=&\mathcal{P}(p)\mathcal{P}(\eta),\\
	\mathcal{P}(p)&=&\prod_k\frac{1}{q_k\Gamma(\alpha_k-1)}\left(\frac{p_k}{q_k}\right)^{-\alpha_k}\exp\left(-\frac{q_k}{p_k}\right),\\
	\mathcal{P}(\eta)&=&\prod_j\frac{1}{r_j\Gamma(\beta_j-1)}\left(\frac{\eta_j}{r_j}\right)^{-\beta_j}\exp\left(-\frac{r_j}{\eta_j}\right).
\end{eqnarray}
The parameters $\alpha_k$ and $\beta_j$ determine the steepness of the power law and the parameters $q_k$ and $r_j$ give the position of the cutoff. In the limit $(\alpha_k,\beta_j)\rightarrow(1,1)$ and $(q_k,r_j)\rightarrow(0,0)$, this turns into the so-called Jeffreys prior, which is flat on a logarithmic scale and can therefore be characterized as non-informative.

\section{Derivation of the filter formulas}
\label{sec:derivation}

With the priors for $s$, $n$, $p$, and $\eta$, we can calculate the joint probability of the signal and the data by marginalizing over the parameters $p$ and $\eta$,
\begin{align}
	\mathcal{P}(s,d)&=\int\mathcal{D}p\int\mathcal{D}\eta~\mathcal{P}(s,d|p,\eta)\mathcal{P}(p,\eta)\nonumber\\
	&=\int\mathcal{D}p\int\mathcal{D}\eta~\mathcal{P}(d|s,p,\eta)\mathcal{P}(s|p)\mathcal{P}(p,\eta)\\
	&=\int\mathcal{D}p\int\mathcal{D}\eta~\mathcal{G}(d-Rs,N)\mathcal{G}(s,S)\mathcal{P}(p,\eta).\nonumber
\end{align}
Solving the integrals yields
\begin{eqnarray}
	\label{eq:joint}
	\mathcal{P}(s,d)=&\prod_k &\frac{\Gamma\left(\gamma_k\right)q_k^{\alpha_k-1}}{\Gamma\left(\alpha_k-1\right)\left(2\pi\right)^{\rho_k/2}}\nonumber\\
	&&\left(q_k+\frac{1}{2}s^\dagger S_k^{-1}s\right)^{-\gamma_k}\nonumber\\
	&\prod_j &\frac{\Gamma\left(\delta_j\right)r_j^{\beta_j-1}}{\Gamma\left(\beta_j-1\right)\left(2\pi\right)^{\mu_j/2}}\\
	&&\left(r_j+\frac{1}{2}\left(d-Rs\right)^\dagger N_j^{-1}\left(d-Rs\right)\right)^{-\delta_j},\nonumber
\end{eqnarray}
where $\rho_k=\mathrm{tr}\left(S_k^{-1}S_k\right)$, $\mu_j=\mathrm{tr}\left(N_j^{-1}N_j\right)$, $\gamma_k=\rho_k/2+\alpha_k-1$, and $\delta_j=\mu_j/2+\beta_j-1$. Note that the posterior is proportional to this joint likelihood for any given dataset.

One could construct the \textit{maximum a posteriori} estimator, however, this was shown in \cite{ensslin_frommert-2011} to perform poorly due to a perception threshold, i.e. modes with too little power in the data are completely filtered out. A better estimate for the posterior mean of the signal can be constructed using the formalism of minimum Gibbs free energy, derived in \cite{ensslin_weig-2010}, where thermodynamic quantities are introduced by identifying the posterior probability density with a canonical density funciton according to
\begin{equation}
	\mathcal{P}(s|d)=\frac{\mathcal{P}(s,d)}{\mathcal{P}(d)}=\frac{\mathrm{e}^{-T^{-1}H(s,d)}}{Z(d)}.
\end{equation}
The Gibbs energy is then
\begin{equation}
	G=U-TS_B,
\end{equation}
where $U=\left<H\right>_{\mathcal{P}(s|d)}$ is the internal energy, $S_B=\left<-\log\mathcal{P}(s|d)\right>_{\mathcal{P}(s|d)}$ the Boltzmann entropy, and $H=-\log\mathcal{P}(s,d)$ is called the information Hamiltonian. The temperature $T$ serves as a tuning parameter, shifting the weight beteween the internal energy term and the entropy term in the Gibbs free energy.

Approximating the posterior with a Gaussian with mean $m$ and covariance $D$,
\begin{equation}
	\label{eq:Gauss-approx}
	\mathcal{P}(s|d)\approx\mathcal{G}(s-m,D),
\end{equation}
gives an approximate internal energy $\tilde{U}$, an approximate entropy $\tilde{S}_B$, and therefore an approximate Gibbs energy
\begin{align}
	\label{eq:Gibbs}
	\tilde{G}(m,D)&=\tilde{U}(m,D)-T\tilde{S}_B\nonumber\\
	&=\left<H(s,d)\right>_{\mathcal{G}(s-m,D)}-\frac{T}{2}\mathrm{tr}\left(1+\log\left(2\pi D\right)\right).
\end{align}
For $T=1$, this approximate energy is, apart from an additive constant, identical to the non-symmetric Kullback-Leibler distance \citep{kullback-1951} between the full posterior and the Gaussian approximation,
\begin{align}
	\label{eq:KL}
	\tilde{G}(m,D)&=\left<H(s,d)+\log\left(\mathcal{G}(s-m,D)\right)\right>_{\mathcal{G}(s-m,D)}\nonumber\\
	&=\int\mathcal{D}s~\mathcal{G}(s-m,D)\log\left(\frac{\mathcal{G}(s-m,D)}{\mathcal{P}(s,d)}\right)\nonumber\\
	&=\int\mathcal{D}s~\mathcal{G}(s-m,D)\log\left(\frac{\mathcal{G}(s-m,D)}{\mathcal{P}(s|d)}\right)\nonumber\\
	&~~~~~+\log\left(\mathcal{P}(d)\right)\nonumber\\
	&=d_\mathrm{KL}\left[\mathcal{G}(s-m,D),\mathcal{P}(s|d)\right]+\log\left(Z(d)\right),
\end{align}
as was shown already in \cite{ensslin_weig-2010}.

The approximate internal energy in our case, calculated from the joint probability of Eq.~\eqref{eq:joint}, is
\begin{align}
	&\tilde{U}(m,D)\cong\sum_k\gamma_k\underbrace{\left<\log\left(q_k+\frac{1}{2}s^\dagger S_k^{-1}s\right)\right>_{\mathcal{G}(s-m,D)}}_{=:\mathcal{A}_k}\nonumber\\
	&+\sum_j\delta_j\underbrace{\left<\log\left(r_j+\frac{1}{2}\left(d-Rs\right)^\dagger N_j^{-1}\left(d-Rs\right)\right)\right>_{\mathcal{G}(s-m,D)}}_{=:\mathcal{B}_j},
\end{align}
where we have dropped terms that are independent of $m$ and $D$. The logarithms can be expanded in an asymptotic power series, giving
\begin{align}
	\label{eq:A_k}
	\mathcal{A}_k=&\log\left(\tilde{q}_k\right)\nonumber\\
	&-\sum_{i=1}^\infty\frac{\left(-1\right)^i}{i\tilde{q}_k^i}\underbrace{\left<\left(q_k+\frac{1}{2}s^\dagger S_k^{-1}s-\tilde{q}_k\right)^i\right>_{\mathcal{G}(s-m,D)}}_{=:\tilde{\mathcal{A}}_ki}
\end{align}
and
\begin{align}
	\mathcal{B}_j&=\log\left(\tilde{r}_j\right)-\sum_{i=1}^\infty\frac{\left(-1\right)^i}{i\tilde{r}_j^i}\nonumber\\
	&\underbrace{\left<\left(r_j+\frac{1}{2}\left(d-Rs\right)^\dagger N_j^{-1}\left(d-Rs\right)-\tilde{r}_j\right)^i\right>_{\mathcal{G}(s-m,D)}}_{=:\tilde{\mathcal{B}}_ji},
	\label{eq:B_j}
\end{align}
where we have chosen the linear dependencies to be captured by
\begin{align}
	\tilde{q}_k&=\left<q_k+\frac{1}{2}s^\dagger S_k^{-1}s\right>_{\mathcal{G}(s-m,D)}\nonumber\\
	&=q_k+\frac{1}{2}\mathrm{tr}\left(\left(mm^\dagger+D\right)S_k^{-1}\right)
\end{align}
and
\begin{align}
	\tilde{r}_j&=\left<r_j+\frac{1}{2}\left(d-Rs\right)^\dagger N_j^{-1}\left(d-Rs\right)\right>_{\mathcal{G}(s-m,D)}\nonumber\\
	&=r_j+\frac{1}{2}\mathrm{tr}\left(\left(\left(d-Rm\right)\left(d-Rm\right)^\dagger+RDR^\dagger\right)N_j^{-1}\right),
\end{align}
respectively.

Here, we restrict ourselves to the zeroth order solution, i.e. we neglect all contributions from $\tilde{\mathcal{A}}$ and $\tilde{\mathcal{B}}$. Furthermore we set $T=1$. The case with $T\neq1$ is discussed up to second order in Appendix \ref{app:higherorder}.

Now we search for the optimal Gaussian approximation to the posterior by minimizing the approximate Gibbs energy, which is equivalent to minimizing the Kullback-Leibler distance between the two probability densities, according to Eq.~\eqref{eq:KL}. Taking the functional derivatives of Eq.~\eqref{eq:Gibbs} with respect to $m$ and $D$ and equating them to zero yields the equations
\begin{eqnarray}
	m&=&Dj,\\
	j&=&\sum_j\frac{\delta_j}{\tilde{r}_j}R^\dagger N_j^{-1}d,\\
	D&=&\left(\sum_k\frac{\gamma_k}{\tilde{q}_k}S_k^{-1}+\sum_j\frac{\delta_j}{\tilde{r}_j}R^\dagger N_j^{-1}R\right)^{-1}.
\end{eqnarray}
By comparing these expressions to the Wiener filter formula, Eq.~\eqref{eq:WF}, we can read off the parameters $p_k=\frac{\tilde{q}_k}{\gamma_k}$ and $\eta_j=\frac{\tilde{r_j}}{\delta_j}$ for the signal and noise covariance matrix, respectively.

So altogether the equations that need to be solved simultaneously are
\begin{subequations}
	\label{eq:gcf}
	\begin{align}
		m&=Dj\\
		p_k&=\frac{q_k+\frac{1}{2}\mathrm{tr}\left(\left(mm^\dagger+D\right)S_k^{-1}\right)}{\frac{\rho_k}{2}+\alpha_k-1}\\
		\eta_j&=\frac{r_j+\frac{1}{2}\mathrm{tr}\left(\left(\left(d-Rm\right)\left(d-Rm\right)^\dagger+RDR^\dagger\right)N_j^{-1}\right)}{\frac{\mu_j}{2}+\beta_j-1}.
	\end{align}
\end{subequations}
Thus, we find both the posterior mean and the posterior covariance for the signal. Note that the first two of these three equations were already found in \cite{ensslin_frommert-2011} and \cite{ensslin_weig-2010}, where the reconstruction of signals with unknown power spectra is discussed. The term \textit{critical filter} was coined in \cite{ensslin_frommert-2011} to refer to this filter since it belongs to a family of filters lying on a line in the parameter plane of \cite{ensslin_frommert-2011} that separates the filters with a perception threshold from those without. The additional uncertainty in the noise covariance that we introduce here simply adds one more equation, leading to an \textit{extended critical filter}.

\section{Application to simulated signals}
\label{sec:application}

\begin{figure*}
	\input{1Da.tex}
	\caption{\label{fig:1Dcomp}Comparison of different filter algorithms in the one-dimensional test case. Each column corresponds to a different setting. The signal, drawn from a power law power spectrum, is the same in each case and depicted in each panel with a solid line. The left column contains homogeneous noise, while in the middle column, the noise is suppressed in the left third of the interval and enhanced in the right third, and in the right column the noise is enhanced in some individual pixels. The first row shows the signal realization along with the data. The second row shows the reconstruction using the Wiener filter formula, assuming the correct power spectrum and under the assumption of homogeneous noise; the third row shows the critical filter reconstruction, assuming the power spectrum to be unknown, but still assuming homogeneous noise. The last row, finally, shows the extended critical filter reconstruction in which both the signal power spectrum as well as the noise variance are assumed to be unknown. The respective reconstructions are depicted by a dashed line which lies on top of the solid one in many cases. In the two right panels of the first row, some of the data points lie outside the area that is shown.}
\end{figure*}
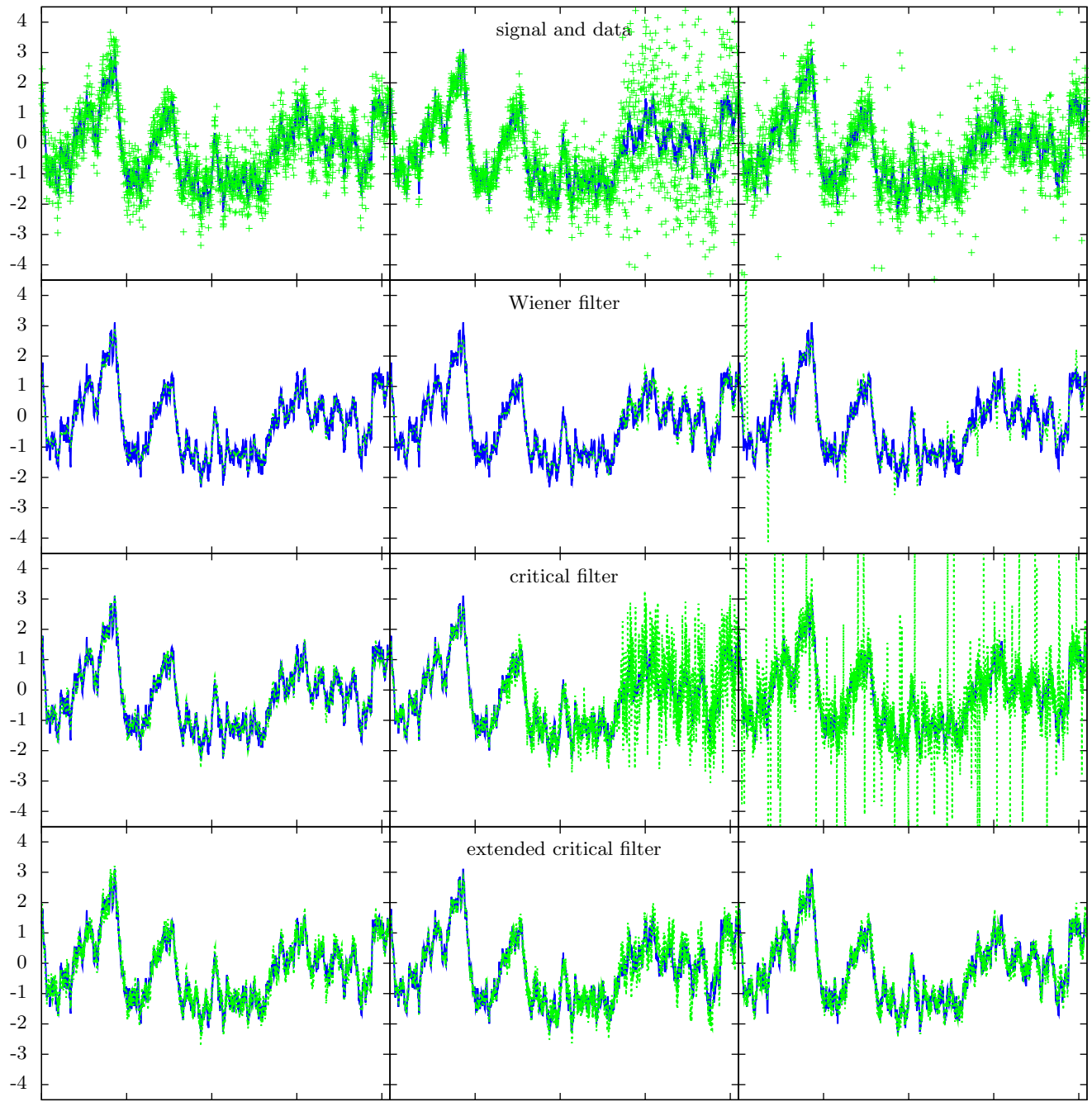

\begin{figure*}
	\includegraphics[width=\textwidth]{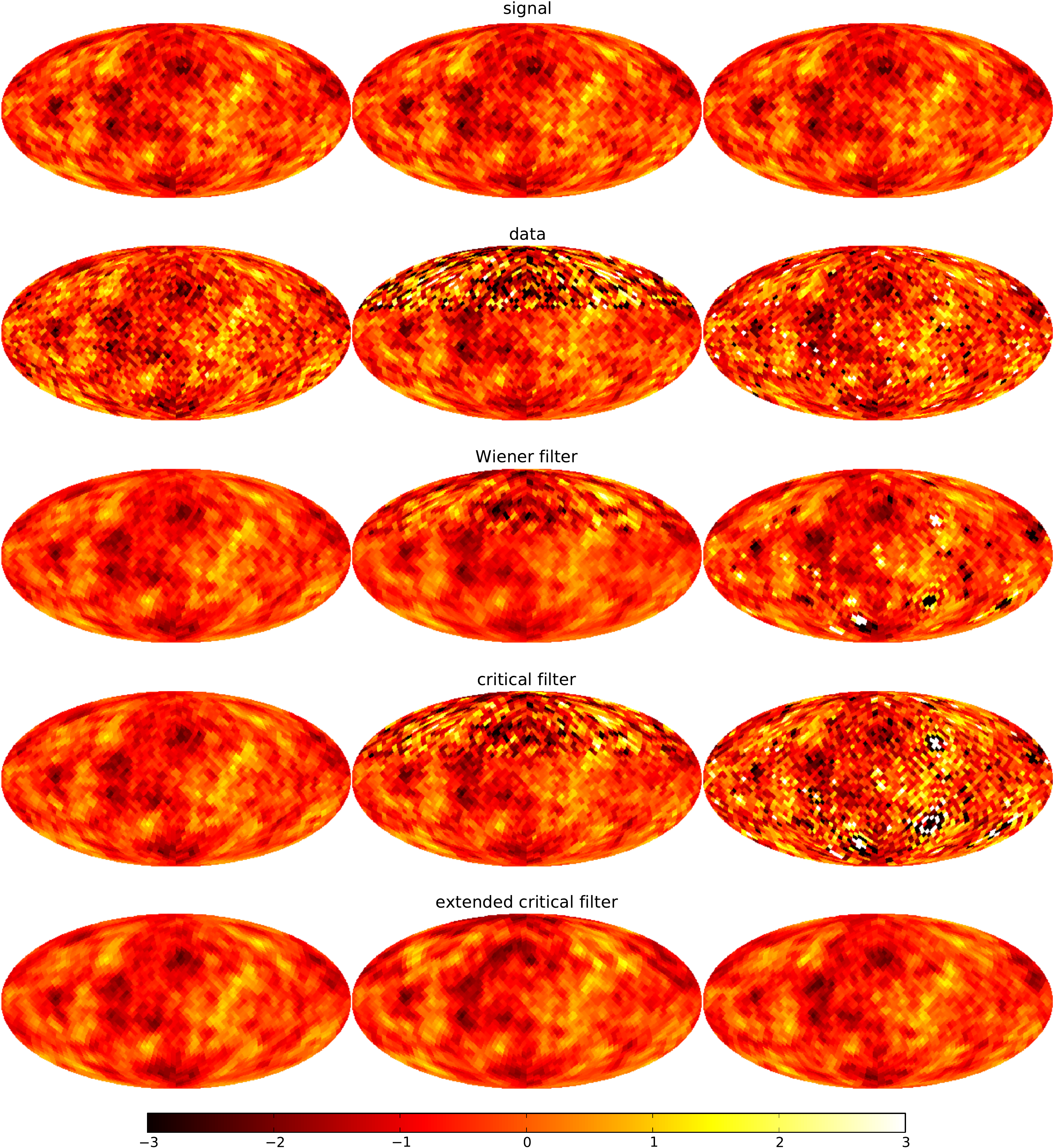}
	\caption{\label{fig:mapcomp}Comparison of different filter algorithms in the spherical case. Each column corresponds to a different setting. The signal, drawn from a power law power spectrum, is the same in each case. The left column contains homogeneous noise, while in the middle column, the noise is suppressed in the southern third of the sphere and enhanced in the northern third, and in the right column the noise is enhanced in some individual pixels. The first row shows the signal realization and the second row the data. The third row shows the reconstruction using the Wiener filter formula, assuming the correct power spectrum and under the assumption of homogeneous noise; the fourth row shows the critical filter reconstruction, assuming the power spectrum to be unknown, but still assuming homogeneous noise. The last row, finally, shows the extended critical filter reconstruction in which both the signal power spectrum as well as the noise variance are assumed to be unknown.}
\end{figure*}

Here we demonstrate the performance of our signal reconstruction algorithm under different circumstances.

\subsection{Setup}

We consider two different scenarios. First, we consider a simple one-dimensional test case, where the signal is supposed to be a real field defined over some interval with periodic boundary conditions. We discretize the interval into $2048$ pixels. For simplicity, we set the response matrix $R$ to be the identity operator, so that our data set consists of $2048$ individual points as well. We further assume statistical homogeneity for the signal field, leading to a covariance matrix that is diagonal in Fourier representation,
\begin{equation}
	S_{kk'}=\left<s_ks_{k'}^*\right>_{\mathcal{P}(s)}=\delta_{kk'}P_s(k),
\end{equation}
with the power spectrum $P_s(k)$ on its diagonal. For this power spectrum we choose a simple power law
\begin{equation}
	P_s(k)\propto\left(1+k\right)^{-2}
\end{equation}
and draw a random realization of the signal from it.

Motivated by astrophysical applications, we also consider a real signal field on the sphere,
\begin{equation}
	s:~\mathcal{S}^2\longrightarrow\mathbb{R}.
\end{equation}
Using again $R=\mathbb{1}$, the data and noise are also fields on the sphere,
\begin{equation}
	d,n:~\mathcal{S}^2\longrightarrow\mathbb{R}.
\end{equation}
In the numerical implementation, we use the \textsc{HEALPix}\footnote{The \textsc{HEALPix} package is available from \url{http://healpix.jpl.nasa.gov}.} discretization scheme at a resolution of $N_{\textrm{side}}=16$, which leads to $3\,072$ pixels. Under the assumption of statistical homogeneity and isotropy, the signal covariance matrix in this case becomes diagonal in the basis given by the spherical harmonics components,
\begin{equation}
	S_{(\ell m)(\ell'm')}=\left<s_{\ell m}s_{\ell'm'}^*\right>_{\mathcal{P}(s)}=\delta_{\ell\ell'}\delta_{mm'}C_\ell,
\end{equation}
where $C_\ell$ are the angular power spectrum components. We draw our signal realization again from a power law spectrum,
\begin{equation}
	C_\ell=\left(1+\ell\right)^{-2}.
\end{equation}

We assume the noise to be uncorrelated in the position basis, making the noise covariance matrix diagonal in this basis,
\begin{equation}
	N_{\hat{n}\hat{n}'}=\left<n_{\hat{n}}n_{\hat{n}'}\right>_{\mathcal{P}(n)}=\delta_{\hat{n}\hat{n}'}\sigma_{\hat{n}}^2,
\end{equation}
where $\hat{n}$ and $\hat{n}'$ denote positions on the sphere or on the interval, respectively. Within this framework, we consider three cases for the noise statistics. In the first one, we use homogeneous noise with variance $\sigma_{\hat{n}}^2=1/4$ independent of $\hat{n}$. For the second case we divide the data space into three zones. In the left/southern third, we suppress the noise variance by a factor of nine and in the right/northern third we enhance it by a factor of nine, while we leave it unchanged in the middle. Finally, in the third case we again assume homogeneous noise with variance $1/4$, but we enhance the variance in five percent of the pixels, randomly selected, by a factor of $100$. Both the signal and the three resulting data realizations are shown in Fig.~\ref{fig:1Dcomp} for the one-dimensional case and in Fig.~\ref{fig:mapcomp} for the spherical case, along with the results of different reconstructions that we discuss next.

\subsection{Reconstructions}

We first apply the standard Wiener filter formula, the results of which are shown in the second row of Fig.~\ref{fig:1Dcomp} for the one-dimensional case and in the middle row of Fig.~\ref{fig:mapcomp} for the spherical case. For this we assume the correct power spectrum to be known, but we assume homogeneous noise with variance $1/4$ in all three cases. In the case where this assumption is correct, the reconstruction is known to be optimal and this is confirmed by visual inspection of the outcome. In the cases with inhomogeneous noise, the Wiener filter fails to completely filter out the noise structures in the data in the regions where the noise is underestimated and therefore reproduces some of them in the reconstruction, as one would expect. This is true for the right (northern) third in the middle column of Fig.~\ref{fig:1Dcomp} (Fig.~\ref{fig:mapcomp}), as well as the noisy pixels in the right column. The opposite should happen in the left (southern) third in the middle column of Fig.~\ref{fig:1Dcomp} (Fig.~\ref{fig:mapcomp}), where the noise is overestimated. One would expect that structures in the data that are actually due to the signal get filtered out. This is actually happening, although it is barely visible in the resulting plots.

Next we assume that the power spectrum is not known a priori, i.e. we apply the critical filter. The resulting plots are shown in the third row of Fig.~\ref{fig:1Dcomp} for the one-dimensional case and the fourth row of Fig.~\ref{fig:mapcomp} for the two-dimensional case. In the one-dimensional case we define the $S_k$ operators of Eq.~\eqref{eq:Sdecomp} to be projections onto bins of width $\Delta_k=2$ in Fourier space, effectively assuming that the two scales that enter the bin have the same power. This binned power is then represented by the parameer $p_k$ in Eq.~\eqref{eq:Sdecomp}. This binning is necessary in one dimension since each individual Fourier component contains only two degrees of freedom.\footnote{Note that since our signal is real, the Fourier components associated with negative $k$-values do not contain additional degrees of freedom.} In the spherical case, we can directly use the angular power spectrum components $C_\ell$ as parameters and the projection-like operators $S_k$ become actual projections onto the $\ell$-th angular scale, which contains $2\ell+1$ degrees of freedom. In both cases we assume Jeffreys prior for the unknown parameters. Then we simply iterate the first two lines of Eq.~\eqref{eq:gcf}, while keeping the assumption of homogeneous noise with variance $1/4$.

In the cases where our assumptions about the noise are true, the resulting map is very close to the Wiener filter reconstruction, confirming the assessment of \cite{ensslin_frommert-2011, ensslin_weig-2010, oppermann-2011a} that the critical filter can yield a very accurate reconstruction, even if the power spectrum is completely unknown. In the cases where we have made false assumptions about the noise, however, we see the same problems that the Wiener filter reconstruction has, only much stronger pronounced. This is because the reconstructed power spectrum now actually accounts for the features in the data that are due to noise where this is underestimated. With this power spectrum, the map reconstruction tends to favor these features even more than when the correct power spectrum is used. This amplifying effect is again much more prominent where the noise was underestimated than where it was overestimated.

Finally, we account for the possibility that we might have misestimated the noise statistics by applying the full extended critical filter, derived in Sect.~\ref{sec:derivation}. As projection-like matrices $N_j$ we choose projections onto the $j$-th pixel of the interval and sphere, respectively, multiplied with our original guess for the noise variance in that pixel, $\sigma_j^2=1/4$. In this way, the parameters $\eta_j$ become correction factors for the noise variance of each data point. For the prior parameters we choose $\beta_j=2$ and we adapt $r_j$ such that $\left<\log\left(\eta_j\right)\right>_{\mathcal{P}(\eta)}=0$. After iterating the full set of equations \eqref{eq:gcf}, we obtain the results shown in the bottom rows of Fig.~\ref{fig:1Dcomp} and Fig.~\ref{fig:mapcomp}. In the case with homogeneous noise, we still get a result that is similar to the Wiener filter one. This shows that we do not lose much by allowing for some uncertainty in the noise covariance. In the cases in which our original noise estimate was wrong, however, we obtain reconstructions of a much higher quality than from the critical filter. Obviously, our algorithm succeeds in uncovering the false error bars in our dataset and correcting them. This works especially well in the case where only individual pixels have underestimated noise variance. This setting makes it especially easy for the algorithm to infer the signal statistics from all the other pixels and find the pixels in which the data points and the signal are inconsistent with one another. However, even in the case where one third of the space is covered with underestimated noise, our algorithm still does a good job in reconstructing the original signal. In the spherical case, the extended critical filter performs even better than the Wiener filter. This is also true for the one-dimensional case in the scenario where the noise in individual pixels is enhanced. In the scenario with enhanced and suppressed noise in one third of the one-dimensional interval, the Wiener filter performs better than the extended critical filter. It should be noted, however, that using the Wiener filter is not an option if the power spectrum of the signal is not known a priori.

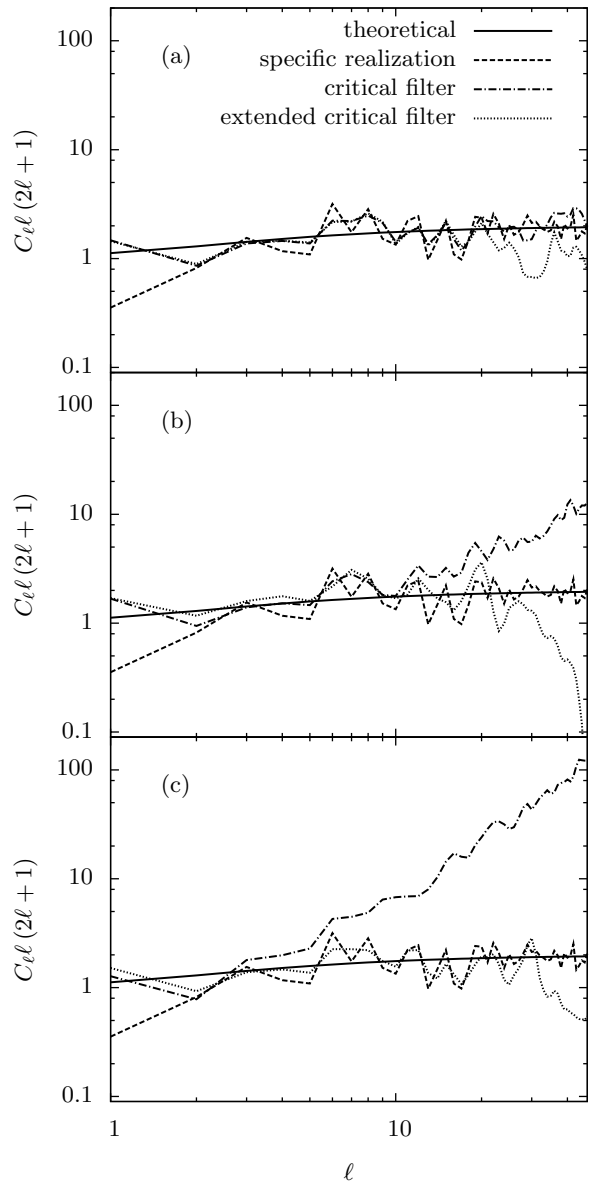
\begin{figure}
	\input{Clsa.tex}
	\caption{\label{fig:Cl}Comparison of the different reconstructed angular power spectra for the spherical scenario. The solid line depicts the theoretical power spectrum which is also used in the Wiener filter reconstructions. The dashed line corresponds to the power of the specific signal realization and the dash-dotted and dotted lines to the power spectrum reconstructed with the critical filter and the extended critical filter, respectively. Panel (a) shows the case with homogeneous noise, panel (b) the one in which the noise is enhanced and suppressed in one third of the sphere each, and panel (c) the one in which the noise is enhanced in individual pixels.}
\end{figure}

\begin{figure*}
	\includegraphics[width=\textwidth]{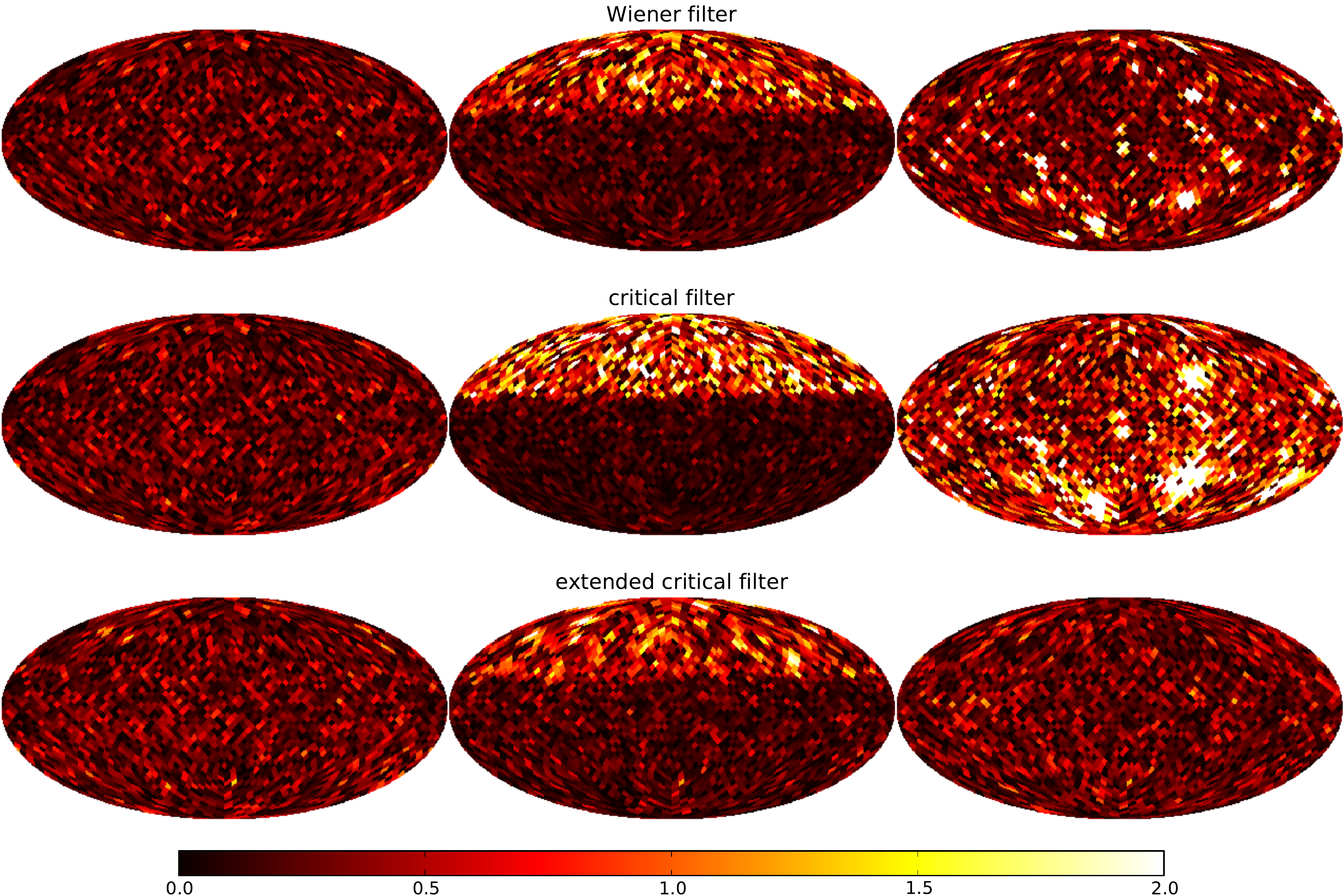}
	\caption{\label{fig:diff}Absolute value of the pixelwise difference between the reconstructed maps and the signal realization for the spherical scenario. Each row shows the results for a different filter algorithm. As in Fig.~\ref{fig:mapcomp}, the left column shows the case with homogeneous noise, the middle column the one with enhanced noise in the northern third of the sphere and suppressed noise in the southern third, and the right column shows the case where the noise is enhanced in individual pixels. Note that the color bar differs from the one used in Fig.~\ref{fig:mapcomp}.}
\end{figure*}

\begin{figure*}
	\includegraphics[width=\textwidth]{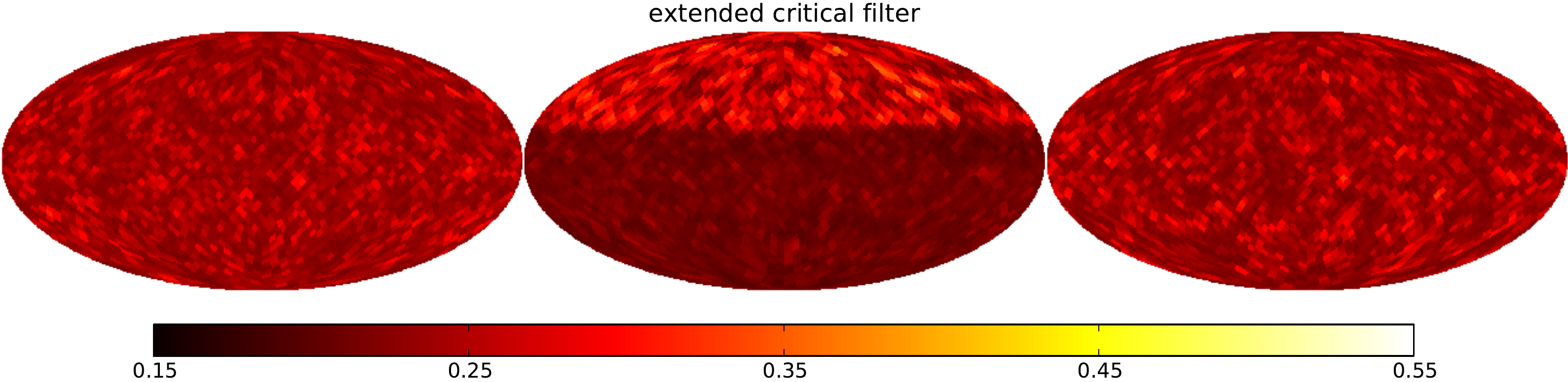}
	\caption{\label{fig:sigma}Pixelwise uncertainty of the extended critical filter reconstructions in the spherical scenario. The left panel shows the case with homogeneous noise, the middle panel the case with enhanced noise in the northern third and suppressed noise in the southern third, and the right panel the case with enhanced noise in individual pixels.}
\end{figure*}

Some further insight can be gained by looking at the reconstructed angular power spectra for the spherical case. These are shown in Fig.~\ref{fig:Cl}. In the case with homogeneous noise, the critical filter recovers the true power spectrum almost perfectly, while the extended critical filter misses some power on the smallest scales, i.e. some of the small-scale power in the data is falsely attributed to noise and therefore not represented in the signal power spectrum. This effect is small, however, and does not greatly influence the resulting map.

In the case in which the noise is highly inhomogeneous, being higher and lower in one third of the data space each, the extended critical filter misses quite a lot of power on small scales. This results in the slightly oversmoothed map seen in Fig.~\ref{fig:mapcomp}. The critical filter, however, that operates under wrong assumptions for the noise statistics, overestimates the power on small scales significantly. This is in agreement with the very noisy reconstructed map.

It is in the third case, in which the noise is greatly enhanced in individual pixels, that the extended critical filter shows its full strength. While the critical filter attributes the power in the faulty pixels to the signal and therefore overestimates the signal power by orders of magnitude, the extended version accounts for the misestimated error bars and does not account for these pixels in the signal power spectrum. While the result is a power spectrum that is slightly underestimated on the smallest scales, this is again only a comparatively small error, still leading to a good reconstruction.

These findings are confirmed by Fig.~\ref{fig:diff}, which shows the differences of the nine reconstructions and the signal realization. Our extension of the critical filter clearly brings the strongest improvement in the case where the noise is enhanced in individual pixels, while also lowering the error in the case with an extended region of underestimated noise. The same can directly be seen for the one-dimensional case in Fig.~\ref{fig:1Dcomp}.

Finally, we plot the standard deviation per pixel of the Gaussian approximation \eqref{eq:Gauss-approx} to the posterior probability distribution, i.e. the square root of the diagonal of the covariance matrix in the pixel basis, $\sigma=\sqrt{\mathrm{diag}(D)}$, in Fig.~\ref{fig:sigma} for the extended critical filter. This can be interpreted as an estimate for the $1\sigma$-error bar of the reconstructed maps. The region with enhanced noise in the second scenario is clearly marked out by a higher uncertainty of the reconstruction due to the corrected entries of the noise covariance matrix entering the information propagator $D$. Note, however, that the full posterior is non-Gaussian and the $1\sigma$ range can therefore not necessarily be interpreted as a $68\%$ confidence interval, especially since we are using a zeroth order approximation to calculate the Gaussian approximation. In fact, in our spherical example only about $50\%$ of the pixels of the reconstructions lie within $1\sigma$ of the correct signal in all three noise scenarios.

\begin{figure*}
	\input{1Ddiffa.tex}
	\caption{\label{fig:1Ddiff}Pixelwise uncertainty of the extended critical filter reconstructions in the one-dimensional scenario. The left panel shows the case with homogeneous noise, the middle panel the case with enhanced noise in the right third and suppressed noise in the left third, and the right panel the case with enhanced noise in individual pixels. The dark curves represent $\pm\sigma$ and the light curve the difference between the reconstruction and the correct signal.}
\end{figure*}
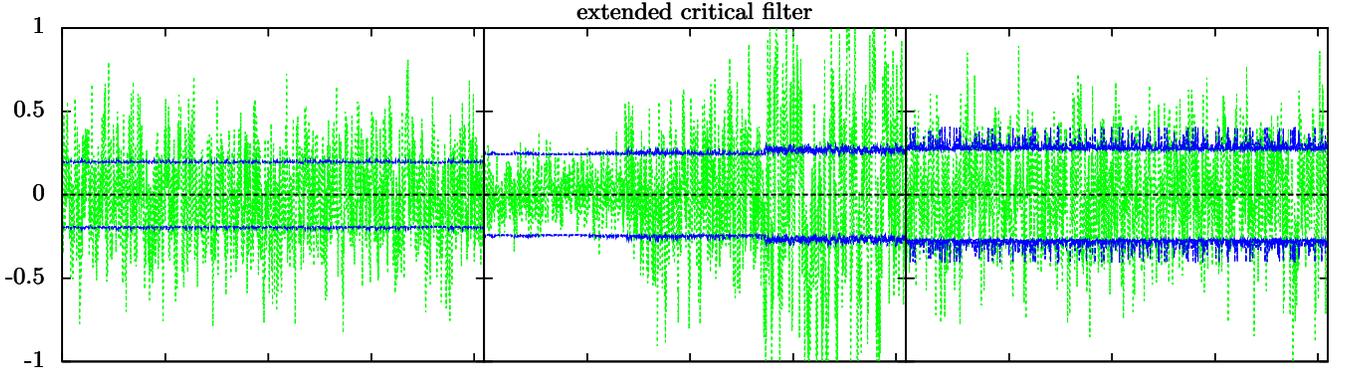

Fig.~\ref{fig:1Ddiff} extends this study to the one-dimensional case. We plot the difference of the signal and the reconstruction result of the extended critical filter, along with lines depicting $\pm\sigma$, for the three different noise settings. In the case where the noise variance is constant within each third of the interval, the one-sigma curve exhibits a step-like behavior at the boundaries of the thirds, although this effect is relatively small. In the case of homogeneous noise, the one-sigma curve is roughly constant while the individual noisy pixels in the last scenario are reflected in the one-sigma curve by its large variations from pixel to pixel. The fraction of the pixels for which the reconstruction lies within $1\sigma$ of the correct signal in the one-dimensional case is 50\% in the case with homogeneous noise, 63\% in the case with enhanced and suppressed noise in a third of the interval each, and 69\% in the case with enhanced noise in individual pixels.

\section{Discussion}
\label{sec:discussion}

Using the formalism of minimum Gibbs free energy we have extended the critical filter algorithm, developed in \cite{ensslin_frommert-2011} and \cite{ensslin_weig-2010}, to an algorithm that allows for uncertainties both in the signal covariance and in the noise covariance. We have demonstrated the performance of our algorithm, Eqs.~\eqref{eq:gcf}, by applying it to a set of mock observations on the sphere, as well as in a simple one-dimensional test case.

These applications have shown that the extended critical filter performs outstandingly if only a few individual data points have a misestimated error bar. However, even in a case where large portions of the data are affected, the algorithm was shown to perform inarguably better than the critical filter, using a fixed -- and faulty -- assumption about the noise statistics. We have also compared the results to those obtained from a Wiener filter reconstruction, using the correct power spectrum, which is known to be optimal if the assumptions about the noise statistics are correct. This filter was demonstrated, however, to lead to reconstructed maps that are much further from the true signal than the results of the extended critical filter in some cases where the assumptions are not correct.

The choice of the two-sphere as the space on which our signal is defined was motivated by astrophysical applications, where we could think of the signal as an all-sky field or some quantity defined on the surface of a star or a planet. Applications in other fields of physics are abundant. However, it should be noted that there is nothing special about the sphere. We could equally well have chosen a more-dimensional euclidean space, using the power spectrum defined in Fourier space instead of the angular power spectrum, as we have done in the one-dimensional scenario.

Furthermore, our choice of the identity operator as response matrix was made only on account of simplicity. It allowed us to represent the data in the same fashion as the signal. It should be clear, however, that the derived filter formulas, Eqs.~\eqref{eq:gcf}, are valid for any response matrix, even a singular one. Applications of the critical filter with non-trivial response matrices were presented in \cite{ensslin_frommert-2011} and \cite{oppermann-2011a} and such a response would not pose a problem for the extended version of the filter.

The problem of signal reconstruction with some uncertainty in the noise variance is certainly one of general interest. There are several ways in which uncertainty in the noise variance might arise. It may be due to questionable assumptions that enter in the calculation of the error bars of the data. Another possibility is that it arises from the definition of the signal itself. The quantity of interest may only be part of what has been measured in the first place in which case the rest of the data would be noise with essentially unknown variance. All these factors come together in the reconstruction problem considered in \cite{oppermann-2011a}. An extension of that reconstruction, using additional sets of data and the improved algorithm presented here, is planned \citep{oppermann-2011c}.

It should be noted, however, that even with the extended critical filter, some knowledge about either the parameters in the signal covariance or the ones in the noise covariance is needed to arrive at a sensible reconstruction. Leaving them both completely free would lead to a degeneracy between signal and noise that cannot be resolved. Only by assigning an informative prior to at least one of the two sets of parameters is this degeneracy broken. Furthermore, the functional bases in which the signal and noise covariances are diagonal, i.e. their eingenspaces, need to differ to allow for a separation of noise from signal.

\begin{acknowledgments}
	The authors would like to thank Henrik Junklewitz and Marco Selig, as well as the anonymous referees, for helpful discussions and comments during the genesis of this work. Some of the results in this paper have been derived using the \textsc{HEALPix} \citep{gorski-2005} package. The computations were performed using the \textsc{Sage} software package \citep{sage}.
\end{acknowledgments}

\appendix

\section{Higher order solutions}
\label{app:higherorder}

Here we briefly list the results for nontrivial temperatures up to second order, i.e. considering terms up to $i=2$ in Eqs.~\eqref{eq:A_k} and \eqref{eq:B_j}. Since the first order terms are zero for our choice of $\tilde{q}_k$ and $\tilde{r}_j$, we list only the resulting filter formulas for the zeroth and second order internal energy.

\subsection{Zeroth order}

The zeroth order solution with arbitrary temperature is rather similar to the one with $T=1$ presented in Sect.~\ref{sec:derivation}. It is given by
\begin{eqnarray}
	m&=&D'j,\\
	j&=&\sum_j\frac{\delta_j}{\tilde{r}_j}R^\dagger N_j^{-1}d,\\
	D'&=&\left(\sum_k\frac{\gamma_k}{\tilde{q}_k}S_k^{-1}+\sum_j\frac{\delta_j}{\tilde{r}_j}R^\dagger N_j^{-1}R\right)^{-1}\\
	D&=&TD'.
\end{eqnarray}
The mean $m$ is completely unchanged. However, the covariance $D$ of the Gaussian approximation is now $T$ times the information propagator $D'$, i.e. the Gaussian approximation becomes wider at higher temperature.

\subsection{Second order}

The second order solution is given by
\begin{widetext}
	\begin{eqnarray}
		m&=&D'j,\\
		j&=&\sum_j\frac{\delta_j}{\tilde{r}_j}Y_jR^\dagger N_j^{-1}d,\\
		D'&=&\left(\sum_k\frac{\gamma_k}{\tilde{q}_k}X_k+\sum_j\frac{\delta_j}{\tilde{r}_j}Y_jR^\dagger N_j^{-1}R\right)^{-1},\\
		X_k&=&1+\frac{1}{\tilde{q}_k^2}\mathrm{tr}\left(\left(mm^\dagger+\frac{1}{2}D\right)S_k^{-1}DS_k^{-1}\right)
		-\frac{1}{\tilde{q}_k}S_k^{-1}D,\\
		Y_j&=&1+\frac{1}{\tilde{r}_j^2}\mathrm{tr}\left(\left(\left(d-Rm\right)\left(d-Rm\right)^\dagger+\frac{1}{2}RDR^\dagger\right)
		N_j^{-1}RDR^\dagger N_j^{-1}\right)-\frac{1}{\tilde{r}_j}R^\dagger N_j^{-1}RD,\\
		D&=&T\left(D'^{-1}-\sum_k\frac{\gamma_k}{\tilde{q}_k^2}S_k^{-1}\left(mm^\dagger\right)S_k^{-1}
		-\sum_j\frac{\delta_j}{\tilde{r}_j^2}R^\dagger N_j^{-1}\left(\left(d-Rm\right)\left(d-Rm\right)^\dagger\right)
		N_j^{-1}R\right)^{-1}.
	\end{eqnarray}
\end{widetext}
Again, the only effect of the temperature is to broaden the approximate Gaussian. However, in the second order solution the operators $X_k$ and $Y_j$ appear, destroying the one-to-one correspondence between the terms in these expressions for $D$, $D'$, and $j$ and the Wiener filter formula Eq.~\eqref{eq:WF}. Therefore, the values of the parameters $p_k$ and $\eta_j$ are not immediately determined by these equations. Note, however, that the goal was not to determine the signal and noise covariance matrices, but to find the optimal Gaussian approximation to the signal posterior, given by $m$ and $D$.

\bibliographystyle{myaa}
\bibliography{noise-rec}

\end{document}

%% file: 1Da.tex
% GNUPLOT: LaTeX picture with Postscript
\begingroup
  \makeatletter
  \providecommand\color[2][]{%
    \GenericError{(gnuplot) \space\space\space\@spaces}{%
      Package color not loaded in conjunction with
      terminal option `colourtext'%
    }{See the gnuplot documentation for explanation.%
    }{Either use 'blacktext' in gnuplot or load the package
      color.sty in LaTeX.}%
    \renewcommand\color[2][]{}%
  }%
  \providecommand\includegraphics[2][]{%
    \GenericError{(gnuplot) \space\space\space\@spaces}{%
      Package graphicx or graphics not loaded%
    }{See the gnuplot documentation for explanation.%
    }{The gnuplot epslatex terminal needs graphicx.sty or graphics.sty.}%
    \renewcommand\includegraphics[2][]{}%
  }%
  \providecommand\rotatebox[2]{#2}%
  \@ifundefined{ifGPcolor}{%
    \newif\ifGPcolor
    \GPcolortrue
  }{}%
  \@ifundefined{ifGPblacktext}{%
    \newif\ifGPblacktext
    \GPblacktexttrue
  }{}%
  % define a \g@addto@macro without @ in the name:
  \let\gplgaddtomacro\g@addto@macro
  % define empty templates for all commands taking text:
  \gdef\gplbacktext{}%
  \gdef\gplfronttext{}%
  \makeatother
  \ifGPblacktext
    % no textcolor at all
    \def\colorrgb#1{}%
    \def\colorgray#1{}%
  \else
    % gray or color?
    \ifGPcolor
      \def\colorrgb#1{\color[rgb]{#1}}%
      \def\colorgray#1{\color[gray]{#1}}%
      \expandafter\def\csname LTw\endcsname{\color{white}}%
      \expandafter\def\csname LTb\endcsname{\color{black}}%
      \expandafter\def\csname LTa\endcsname{\color{black}}%
      \expandafter\def\csname LT0\endcsname{\color[rgb]{1,0,0}}%
      \expandafter\def\csname LT1\endcsname{\color[rgb]{0,1,0}}%
      \expandafter\def\csname LT2\endcsname{\color[rgb]{0,0,1}}%
      \expandafter\def\csname LT3\endcsname{\color[rgb]{1,0,1}}%
      \expandafter\def\csname LT4\endcsname{\color[rgb]{0,1,1}}%
      \expandafter\def\csname LT5\endcsname{\color[rgb]{1,1,0}}%
      \expandafter\def\csname LT6\endcsname{\color[rgb]{0,0,0}}%
      \expandafter\def\csname LT7\endcsname{\color[rgb]{1,0.3,0}}%
      \expandafter\def\csname LT8\endcsname{\color[rgb]{0.5,0.5,0.5}}%
    \else
      % gray
      \def\colorrgb#1{\color{black}}%
      \def\colorgray#1{\color[gray]{#1}}%
      \expandafter\def\csname LTw\endcsname{\color{white}}%
      \expandafter\def\csname LTb\endcsname{\color{black}}%
      \expandafter\def\csname LTa\endcsname{\color{black}}%
      \expandafter\def\csname LT0\endcsname{\color{black}}%
      \expandafter\def\csname LT1\endcsname{\color{black}}%
      \expandafter\def\csname LT2\endcsname{\color{black}}%
      \expandafter\def\csname LT3\endcsname{\color{black}}%
      \expandafter\def\csname LT4\endcsname{\color{black}}%
      \expandafter\def\csname LT5\endcsname{\color{black}}%
      \expandafter\def\csname LT6\endcsname{\color{black}}%
      \expandafter\def\csname LT7\endcsname{\color{black}}%
      \expandafter\def\csname LT8\endcsname{\color{black}}%
    \fi
  \fi
  \setlength{\unitlength}{0.0500bp}%
  \begin{picture}(9936.00,10080.00)%
    \gplgaddtomacro\gplbacktext{%
      \csname LTb\endcsname%
      \put(215,7674){\makebox(0,0)[r]{\strut{}-4}}%
      \put(215,7951){\makebox(0,0)[r]{\strut{}-3}}%
      \put(215,8228){\makebox(0,0)[r]{\strut{}-2}}%
      \put(215,8505){\makebox(0,0)[r]{\strut{}-1}}%
      \put(215,8782){\makebox(0,0)[r]{\strut{} 0}}%
      \put(215,9059){\makebox(0,0)[r]{\strut{} 1}}%
      \put(215,9336){\makebox(0,0)[r]{\strut{} 2}}%
      \put(215,9613){\makebox(0,0)[r]{\strut{} 3}}%
      \put(215,9890){\makebox(0,0)[r]{\strut{} 4}}%
      \put(347,7315){\makebox(0,0){\strut{}}}%
      \put(1123,7315){\makebox(0,0){\strut{}}}%
      \put(1900,7315){\makebox(0,0){\strut{}}}%
      \put(2676,7315){\makebox(0,0){\strut{}}}%
      \put(3452,7315){\makebox(0,0){\strut{}}}%
    }%
    \gplgaddtomacro\gplfronttext{%
    }%
    \gplgaddtomacro\gplbacktext{%
      \csname LTb\endcsname%
      \put(215,5179){\makebox(0,0)[r]{\strut{}-4}}%
      \put(215,5456){\makebox(0,0)[r]{\strut{}-3}}%
      \put(215,5733){\makebox(0,0)[r]{\strut{}-2}}%
      \put(215,6010){\makebox(0,0)[r]{\strut{}-1}}%
      \put(215,6287){\makebox(0,0)[r]{\strut{} 0}}%
      \put(215,6564){\makebox(0,0)[r]{\strut{} 1}}%
      \put(215,6841){\makebox(0,0)[r]{\strut{} 2}}%
      \put(215,7118){\makebox(0,0)[r]{\strut{} 3}}%
      \put(215,7395){\makebox(0,0)[r]{\strut{} 4}}%
      \put(347,4820){\makebox(0,0){\strut{}}}%
      \put(1123,4820){\makebox(0,0){\strut{}}}%
      \put(1900,4820){\makebox(0,0){\strut{}}}%
      \put(2676,4820){\makebox(0,0){\strut{}}}%
      \put(3452,4820){\makebox(0,0){\strut{}}}%
    }%
    \gplgaddtomacro\gplfronttext{%
    }%
    \gplgaddtomacro\gplbacktext{%
      \csname LTb\endcsname%
      \put(215,2684){\makebox(0,0)[r]{\strut{}-4}}%
      \put(215,2961){\makebox(0,0)[r]{\strut{}-3}}%
      \put(215,3238){\makebox(0,0)[r]{\strut{}-2}}%
      \put(215,3515){\makebox(0,0)[r]{\strut{}-1}}%
      \put(215,3792){\makebox(0,0)[r]{\strut{} 0}}%
      \put(215,4069){\makebox(0,0)[r]{\strut{} 1}}%
      \put(215,4346){\makebox(0,0)[r]{\strut{} 2}}%
      \put(215,4623){\makebox(0,0)[r]{\strut{} 3}}%
      \put(215,4900){\makebox(0,0)[r]{\strut{} 4}}%
      \put(347,2325){\makebox(0,0){\strut{}}}%
      \put(1123,2325){\makebox(0,0){\strut{}}}%
      \put(1900,2325){\makebox(0,0){\strut{}}}%
      \put(2676,2325){\makebox(0,0){\strut{}}}%
      \put(3452,2325){\makebox(0,0){\strut{}}}%
    }%
    \gplgaddtomacro\gplfronttext{%
    }%
    \gplgaddtomacro\gplbacktext{%
      \csname LTb\endcsname%
      \put(215,189){\makebox(0,0)[r]{\strut{}-4}}%
      \put(215,466){\makebox(0,0)[r]{\strut{}-3}}%
      \put(215,743){\makebox(0,0)[r]{\strut{}-2}}%
      \put(215,1020){\makebox(0,0)[r]{\strut{}-1}}%
      \put(215,1297){\makebox(0,0)[r]{\strut{} 0}}%
      \put(215,1574){\makebox(0,0)[r]{\strut{} 1}}%
      \put(215,1851){\makebox(0,0)[r]{\strut{} 2}}%
      \put(215,2128){\makebox(0,0)[r]{\strut{} 3}}%
      \put(215,2405){\makebox(0,0)[r]{\strut{} 4}}%
      \put(347,-170){\makebox(0,0){\strut{}}}%
      \put(1123,-170){\makebox(0,0){\strut{}}}%
      \put(1900,-170){\makebox(0,0){\strut{}}}%
      \put(2676,-170){\makebox(0,0){\strut{}}}%
      \put(3452,-170){\makebox(0,0){\strut{}}}%
    }%
    \gplgaddtomacro\gplfronttext{%
    }%
    \gplgaddtomacro\gplbacktext{%
      \csname LTb\endcsname%
      \put(3395,7674){\makebox(0,0)[r]{\strut{}}}%
      \put(3395,7951){\makebox(0,0)[r]{\strut{}}}%
      \put(3395,8228){\makebox(0,0)[r]{\strut{}}}%
      \put(3395,8505){\makebox(0,0)[r]{\strut{}}}%
      \put(3395,8782){\makebox(0,0)[r]{\strut{}}}%
      \put(3395,9059){\makebox(0,0)[r]{\strut{}}}%
      \put(3395,9336){\makebox(0,0)[r]{\strut{}}}%
      \put(3395,9613){\makebox(0,0)[r]{\strut{}}}%
      \put(3395,9890){\makebox(0,0)[r]{\strut{}}}%
      \put(3527,7315){\makebox(0,0){\strut{}}}%
      \put(4303,7315){\makebox(0,0){\strut{}}}%
      \put(5079,7315){\makebox(0,0){\strut{}}}%
      \put(5855,7315){\makebox(0,0){\strut{}}}%
      \put(6631,7315){\makebox(0,0){\strut{}}}%
      \put(5116,9809){\makebox(0,0){\strut{}signal and data}}%
    }%
    \gplgaddtomacro\gplfronttext{%
    }%
    \gplgaddtomacro\gplbacktext{%
      \csname LTb\endcsname%
      \put(3395,5179){\makebox(0,0)[r]{\strut{}}}%
      \put(3395,5456){\makebox(0,0)[r]{\strut{}}}%
      \put(3395,5733){\makebox(0,0)[r]{\strut{}}}%
      \put(3395,6010){\makebox(0,0)[r]{\strut{}}}%
      \put(3395,6287){\makebox(0,0)[r]{\strut{}}}%
      \put(3395,6564){\makebox(0,0)[r]{\strut{}}}%
      \put(3395,6841){\makebox(0,0)[r]{\strut{}}}%
      \put(3395,7118){\makebox(0,0)[r]{\strut{}}}%
      \put(3395,7395){\makebox(0,0)[r]{\strut{}}}%
      \put(3527,4820){\makebox(0,0){\strut{}}}%
      \put(4303,4820){\makebox(0,0){\strut{}}}%
      \put(5079,4820){\makebox(0,0){\strut{}}}%
      \put(5855,4820){\makebox(0,0){\strut{}}}%
      \put(6631,4820){\makebox(0,0){\strut{}}}%
      \put(5116,7314){\makebox(0,0){\strut{}Wiener filter}}%
    }%
    \gplgaddtomacro\gplfronttext{%
    }%
    \gplgaddtomacro\gplbacktext{%
      \csname LTb\endcsname%
      \put(3395,2684){\makebox(0,0)[r]{\strut{}}}%
      \put(3395,2961){\makebox(0,0)[r]{\strut{}}}%
      \put(3395,3238){\makebox(0,0)[r]{\strut{}}}%
      \put(3395,3515){\makebox(0,0)[r]{\strut{}}}%
      \put(3395,3792){\makebox(0,0)[r]{\strut{}}}%
      \put(3395,4069){\makebox(0,0)[r]{\strut{}}}%
      \put(3395,4346){\makebox(0,0)[r]{\strut{}}}%
      \put(3395,4623){\makebox(0,0)[r]{\strut{}}}%
      \put(3395,4900){\makebox(0,0)[r]{\strut{}}}%
      \put(3527,2325){\makebox(0,0){\strut{}}}%
      \put(4303,2325){\makebox(0,0){\strut{}}}%
      \put(5079,2325){\makebox(0,0){\strut{}}}%
      \put(5855,2325){\makebox(0,0){\strut{}}}%
      \put(6631,2325){\makebox(0,0){\strut{}}}%
      \put(5116,4819){\makebox(0,0){\strut{}critical filter}}%
    }%
    \gplgaddtomacro\gplfronttext{%
    }%
    \gplgaddtomacro\gplbacktext{%
      \csname LTb\endcsname%
      \put(3395,189){\makebox(0,0)[r]{\strut{}}}%
      \put(3395,466){\makebox(0,0)[r]{\strut{}}}%
      \put(3395,743){\makebox(0,0)[r]{\strut{}}}%
      \put(3395,1020){\makebox(0,0)[r]{\strut{}}}%
      \put(3395,1297){\makebox(0,0)[r]{\strut{}}}%
      \put(3395,1574){\makebox(0,0)[r]{\strut{}}}%
      \put(3395,1851){\makebox(0,0)[r]{\strut{}}}%
      \put(3395,2128){\makebox(0,0)[r]{\strut{}}}%
      \put(3395,2405){\makebox(0,0)[r]{\strut{}}}%
      \put(3527,-170){\makebox(0,0){\strut{}}}%
      \put(4303,-170){\makebox(0,0){\strut{}}}%
      \put(5079,-170){\makebox(0,0){\strut{}}}%
      \put(5855,-170){\makebox(0,0){\strut{}}}%
      \put(6631,-170){\makebox(0,0){\strut{}}}%
      \put(5116,2324){\makebox(0,0){\strut{}extended critical filter}}%
    }%
    \gplgaddtomacro\gplfronttext{%
    }%
    \gplgaddtomacro\gplbacktext{%
      \csname LTb\endcsname%
      \put(6574,7674){\makebox(0,0)[r]{\strut{}}}%
      \put(6574,7951){\makebox(0,0)[r]{\strut{}}}%
      \put(6574,8228){\makebox(0,0)[r]{\strut{}}}%
      \put(6574,8505){\makebox(0,0)[r]{\strut{}}}%
      \put(6574,8782){\makebox(0,0)[r]{\strut{}}}%
      \put(6574,9059){\makebox(0,0)[r]{\strut{}}}%
      \put(6574,9336){\makebox(0,0)[r]{\strut{}}}%
      \put(6574,9613){\makebox(0,0)[r]{\strut{}}}%
      \put(6574,9890){\makebox(0,0)[r]{\strut{}}}%
      \put(6706,7315){\makebox(0,0){\strut{}}}%
      \put(7482,7315){\makebox(0,0){\strut{}}}%
      \put(8259,7315){\makebox(0,0){\strut{}}}%
      \put(9035,7315){\makebox(0,0){\strut{}}}%
      \put(9811,7315){\makebox(0,0){\strut{}}}%
      \put(8296,9919){\makebox(0,0){\strut{}}}%
    }%
    \gplgaddtomacro\gplfronttext{%
    }%
    \gplgaddtomacro\gplbacktext{%
      \csname LTb\endcsname%
      \put(6574,5179){\makebox(0,0)[r]{\strut{}}}%
      \put(6574,5456){\makebox(0,0)[r]{\strut{}}}%
      \put(6574,5733){\makebox(0,0)[r]{\strut{}}}%
      \put(6574,6010){\makebox(0,0)[r]{\strut{}}}%
      \put(6574,6287){\makebox(0,0)[r]{\strut{}}}%
      \put(6574,6564){\makebox(0,0)[r]{\strut{}}}%
      \put(6574,6841){\makebox(0,0)[r]{\strut{}}}%
      \put(6574,7118){\makebox(0,0)[r]{\strut{}}}%
      \put(6574,7395){\makebox(0,0)[r]{\strut{}}}%
      \put(6706,4820){\makebox(0,0){\strut{}}}%
      \put(7482,4820){\makebox(0,0){\strut{}}}%
      \put(8259,4820){\makebox(0,0){\strut{}}}%
      \put(9035,4820){\makebox(0,0){\strut{}}}%
      \put(9811,4820){\makebox(0,0){\strut{}}}%
      \put(8296,7424){\makebox(0,0){\strut{}}}%
    }%
    \gplgaddtomacro\gplfronttext{%
    }%
    \gplgaddtomacro\gplbacktext{%
      \csname LTb\endcsname%
      \put(6574,2684){\makebox(0,0)[r]{\strut{}}}%
      \put(6574,2961){\makebox(0,0)[r]{\strut{}}}%
      \put(6574,3238){\makebox(0,0)[r]{\strut{}}}%
      \put(6574,3515){\makebox(0,0)[r]{\strut{}}}%
      \put(6574,3792){\makebox(0,0)[r]{\strut{}}}%
      \put(6574,4069){\makebox(0,0)[r]{\strut{}}}%
      \put(6574,4346){\makebox(0,0)[r]{\strut{}}}%
      \put(6574,4623){\makebox(0,0)[r]{\strut{}}}%
      \put(6574,4900){\makebox(0,0)[r]{\strut{}}}%
      \put(6706,2325){\makebox(0,0){\strut{}}}%
      \put(7482,2325){\makebox(0,0){\strut{}}}%
      \put(8259,2325){\makebox(0,0){\strut{}}}%
      \put(9035,2325){\makebox(0,0){\strut{}}}%
      \put(9811,2325){\makebox(0,0){\strut{}}}%
      \put(8296,4929){\makebox(0,0){\strut{}}}%
    }%
    \gplgaddtomacro\gplfronttext{%
    }%
    \gplgaddtomacro\gplbacktext{%
      \csname LTb\endcsname%
      \put(6574,189){\makebox(0,0)[r]{\strut{}}}%
      \put(6574,466){\makebox(0,0)[r]{\strut{}}}%
      \put(6574,743){\makebox(0,0)[r]{\strut{}}}%
      \put(6574,1020){\makebox(0,0)[r]{\strut{}}}%
      \put(6574,1297){\makebox(0,0)[r]{\strut{}}}%
      \put(6574,1574){\makebox(0,0)[r]{\strut{}}}%
      \put(6574,1851){\makebox(0,0)[r]{\strut{}}}%
      \put(6574,2128){\makebox(0,0)[r]{\strut{}}}%
      \put(6574,2405){\makebox(0,0)[r]{\strut{}}}%
      \put(6706,-170){\makebox(0,0){\strut{}}}%
      \put(7482,-170){\makebox(0,0){\strut{}}}%
      \put(8259,-170){\makebox(0,0){\strut{}}}%
      \put(9035,-170){\makebox(0,0){\strut{}}}%
      \put(9811,-170){\makebox(0,0){\strut{}}}%
      \put(8296,2434){\makebox(0,0){\strut{}}}%
    }%
    \gplgaddtomacro\gplfronttext{%
    }%
    \gplbacktext
    \put(0,0){\includegraphics{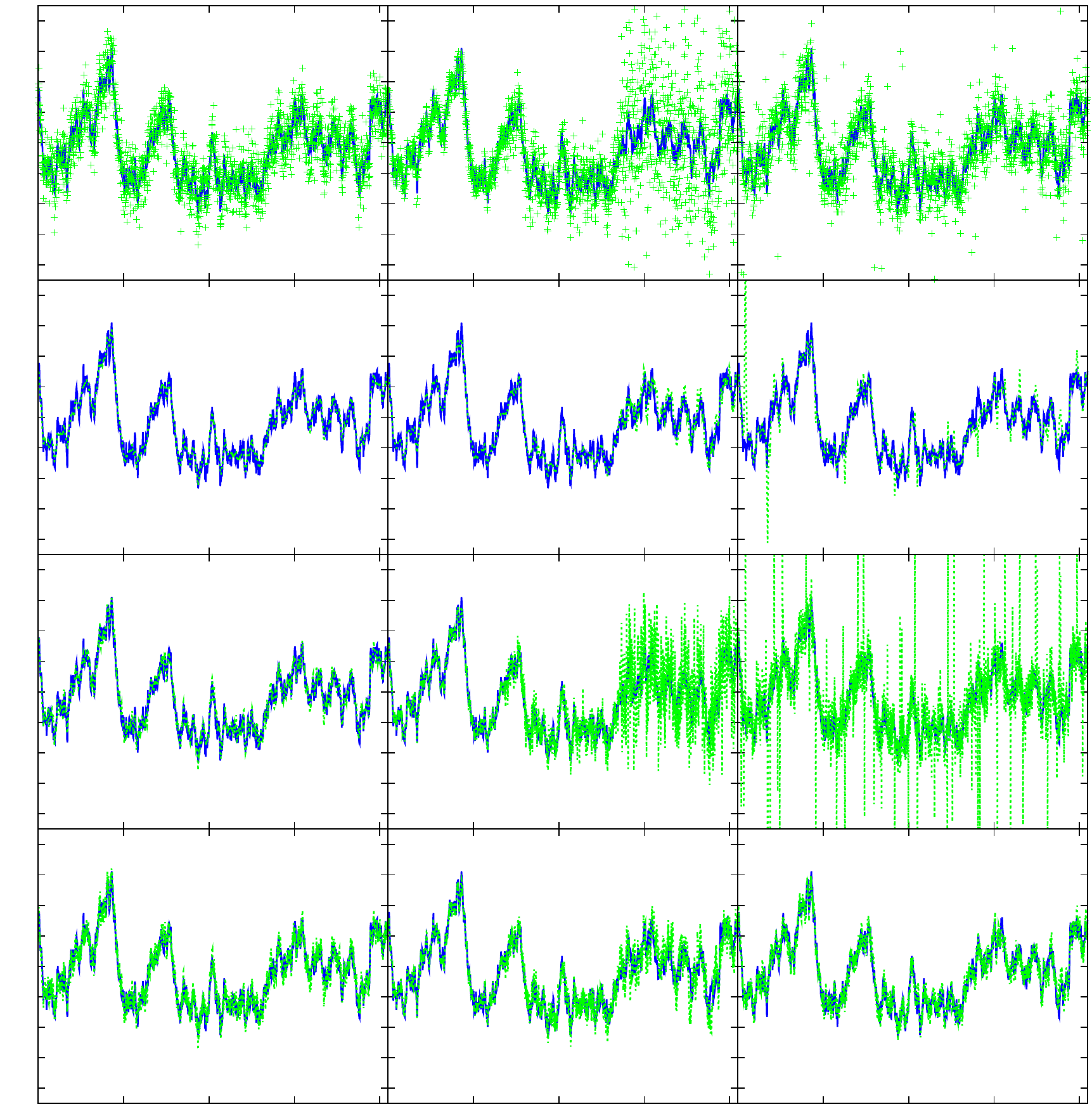}}%
    \gplfronttext
  \end{picture}%
\endgroup

%% file: Clsa.tex
% GNUPLOT: LaTeX picture with Postscript
\begingroup
  \makeatletter
  \providecommand\color[2][]{%
    \GenericError{(gnuplot) \space\space\space\@spaces}{%
      Package color not loaded in conjunction with
      terminal option `colourtext'%
    }{See the gnuplot documentation for explanation.%
    }{Either use 'blacktext' in gnuplot or load the package
      color.sty in LaTeX.}%
    \renewcommand\color[2][]{}%
  }%
  \providecommand\includegraphics[2][]{%
    \GenericError{(gnuplot) \space\space\space\@spaces}{%
      Package graphicx or graphics not loaded%
    }{See the gnuplot documentation for explanation.%
    }{The gnuplot epslatex terminal needs graphicx.sty or graphics.sty.}%
    \renewcommand\includegraphics[2][]{}%
  }%
  \providecommand\rotatebox[2]{#2}%
  \@ifundefined{ifGPcolor}{%
    \newif\ifGPcolor
    \GPcolorfalse
  }{}%
  \@ifundefined{ifGPblacktext}{%
    \newif\ifGPblacktext
    \GPblacktexttrue
  }{}%
  % define a \g@addto@macro without @ in the name:
  \let\gplgaddtomacro\g@addto@macro
  % define empty templates for all commands taking text:
  \gdef\gplbacktext{}%
  \gdef\gplfronttext{}%
  \makeatother
  \ifGPblacktext
    % no textcolor at all
    \def\colorrgb#1{}%
    \def\colorgray#1{}%
  \else
    % gray or color?
    \ifGPcolor
      \def\colorrgb#1{\color[rgb]{#1}}%
      \def\colorgray#1{\color[gray]{#1}}%
      \expandafter\def\csname LTw\endcsname{\color{white}}%
      \expandafter\def\csname LTb\endcsname{\color{black}}%
      \expandafter\def\csname LTa\endcsname{\color{black}}%
      \expandafter\def\csname LT0\endcsname{\color[rgb]{1,0,0}}%
      \expandafter\def\csname LT1\endcsname{\color[rgb]{0,1,0}}%
      \expandafter\def\csname LT2\endcsname{\color[rgb]{0,0,1}}%
      \expandafter\def\csname LT3\endcsname{\color[rgb]{1,0,1}}%
      \expandafter\def\csname LT4\endcsname{\color[rgb]{0,1,1}}%
      \expandafter\def\csname LT5\endcsname{\color[rgb]{1,1,0}}%
      \expandafter\def\csname LT6\endcsname{\color[rgb]{0,0,0}}%
      \expandafter\def\csname LT7\endcsname{\color[rgb]{1,0.3,0}}%
      \expandafter\def\csname LT8\endcsname{\color[rgb]{0.5,0.5,0.5}}%
    \else
      % gray
      \def\colorrgb#1{\color{black}}%
      \def\colorgray#1{\color[gray]{#1}}%
      \expandafter\def\csname LTw\endcsname{\color{white}}%
      \expandafter\def\csname LTb\endcsname{\color{black}}%
      \expandafter\def\csname LTa\endcsname{\color{black}}%
      \expandafter\def\csname LT0\endcsname{\color{black}}%
      \expandafter\def\csname LT1\endcsname{\color{black}}%
      \expandafter\def\csname LT2\endcsname{\color{black}}%
      \expandafter\def\csname LT3\endcsname{\color{black}}%
      \expandafter\def\csname LT4\endcsname{\color{black}}%
      \expandafter\def\csname LT5\endcsname{\color{black}}%
      \expandafter\def\csname LT6\endcsname{\color{black}}%
      \expandafter\def\csname LT7\endcsname{\color{black}}%
      \expandafter\def\csname LT8\endcsname{\color{black}}%
    \fi
  \fi
  \setlength{\unitlength}{0.0500bp}%
  \begin{picture}(4030.00,8870.00)%
    \gplgaddtomacro\gplbacktext{%
      \csname LTb\endcsname%
      \put(264,6114){\makebox(0,0)[r]{\strut{} 0.1}}%
      \put(264,6935){\makebox(0,0)[r]{\strut{} 1}}%
      \put(264,7756){\makebox(0,0)[r]{\strut{} 10}}%
      \put(264,8578){\makebox(0,0)[r]{\strut{} 100}}%
      \put(396,5856){\makebox(0,0){\strut{}}}%
      \put(2545,5856){\makebox(0,0){\strut{}}}%
      \put(-242,7450){\rotatebox{90}{\makebox(0,0){\strut{}$C_\ell\ell\left(2\ell+1\right)$}}}%
      \put(774,8451){\makebox(0,0)[l]{\strut{}(a)}}%
    }%
    \gplgaddtomacro\gplfronttext{%
      \csname LTb\endcsname%
      \put(3002,8652){\makebox(0,0)[r]{\strut{}theoretical}}%
      \csname LTb\endcsname%
      \put(3002,8432){\makebox(0,0)[r]{\strut{}specific realization}}%
      \csname LTb\endcsname%
      \put(3002,8212){\makebox(0,0)[r]{\strut{}critical filter}}%
      \csname LTb\endcsname%
      \put(3002,7992){\makebox(0,0)[r]{\strut{}extended critical filter}}%
    }%
    \gplgaddtomacro\gplbacktext{%
      \csname LTb\endcsname%
      \put(264,3364){\makebox(0,0)[r]{\strut{} 0.1}}%
      \put(264,4185){\makebox(0,0)[r]{\strut{} 1}}%
      \put(264,5006){\makebox(0,0)[r]{\strut{} 10}}%
      \put(264,5828){\makebox(0,0)[r]{\strut{} 100}}%
      \put(396,3106){\makebox(0,0){\strut{}}}%
      \put(2545,3106){\makebox(0,0){\strut{}}}%
      \put(-242,4700){\rotatebox{90}{\makebox(0,0){\strut{}$C_\ell\ell\left(2\ell+1\right)$}}}%
      \put(774,5701){\makebox(0,0)[l]{\strut{}(b)}}%
    }%
    \gplgaddtomacro\gplfronttext{%
    }%
    \gplgaddtomacro\gplbacktext{%
      \csname LTb\endcsname%
      \put(264,615){\makebox(0,0)[r]{\strut{} 0.1}}%
      \put(264,1436){\makebox(0,0)[r]{\strut{} 1}}%
      \put(264,2257){\makebox(0,0)[r]{\strut{} 10}}%
      \put(264,3078){\makebox(0,0)[r]{\strut{} 100}}%
      \put(396,357){\makebox(0,0){\strut{} 1}}%
      \put(2545,357){\makebox(0,0){\strut{} 10}}%
      \put(-242,1951){\rotatebox{90}{\makebox(0,0){\strut{}$C_\ell\ell\left(2\ell+1\right)$}}}%
      \put(2192,27){\makebox(0,0){\strut{}$\ell$}}%
      \put(774,2951){\makebox(0,0)[l]{\strut{}(c)}}%
    }%
    \gplgaddtomacro\gplfronttext{%
    }%
    \gplbacktext
    \put(0,0){\includegraphics{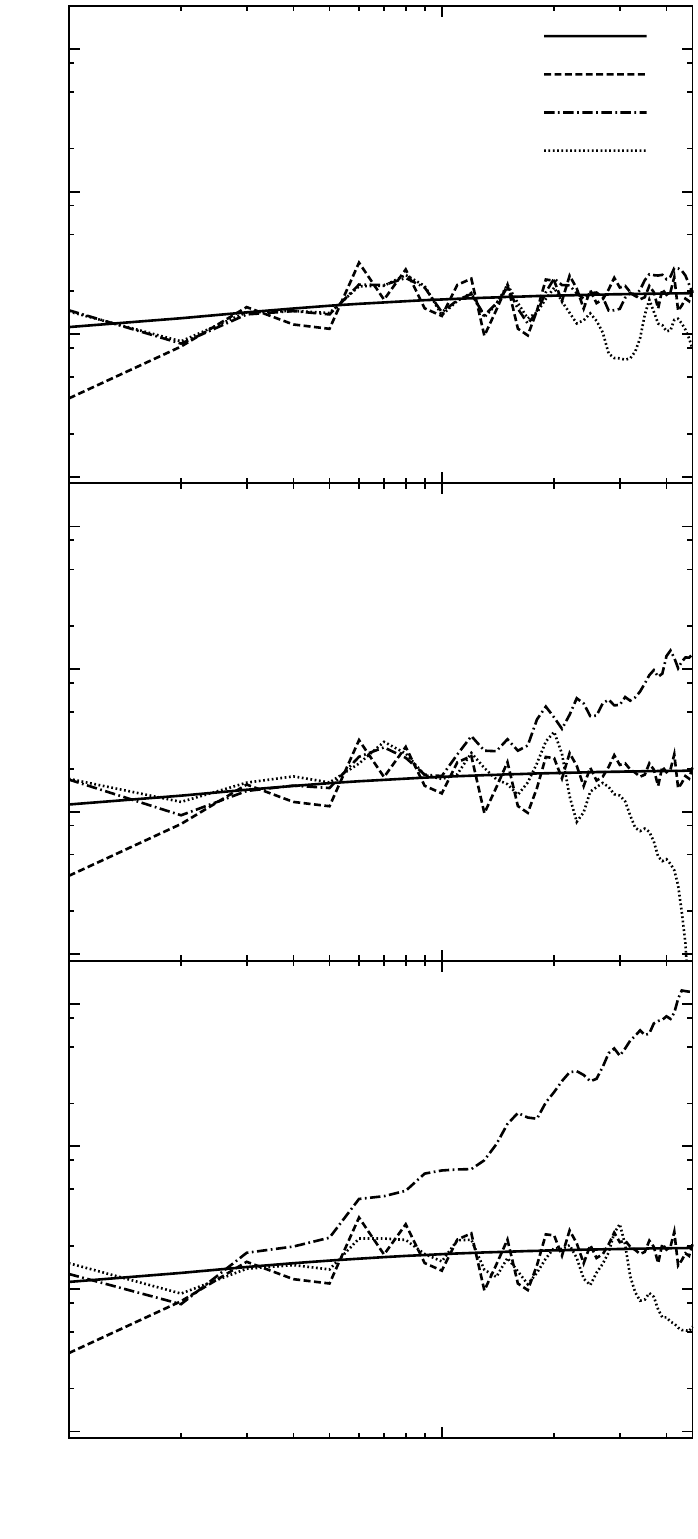}}%
    \gplfronttext
  \end{picture}%
\endgroup

%% file: 1Ddiffa.tex
% GNUPLOT: LaTeX picture with Postscript
\begingroup
  \makeatletter
  \providecommand\color[2][]{%
    \GenericError{(gnuplot) \space\space\space\@spaces}{%
      Package color not loaded in conjunction with
      terminal option `colourtext'%
    }{See the gnuplot documentation for explanation.%
    }{Either use 'blacktext' in gnuplot or load the package
      color.sty in LaTeX.}%
    \renewcommand\color[2][]{}%
  }%
  \providecommand\includegraphics[2][]{%
    \GenericError{(gnuplot) \space\space\space\@spaces}{%
      Package graphicx or graphics not loaded%
    }{See the gnuplot documentation for explanation.%
    }{The gnuplot epslatex terminal needs graphicx.sty or graphics.sty.}%
    \renewcommand\includegraphics[2][]{}%
  }%
  \providecommand\rotatebox[2]{#2}%
  \@ifundefined{ifGPcolor}{%
    \newif\ifGPcolor
    \GPcolortrue
  }{}%
  \@ifundefined{ifGPblacktext}{%
    \newif\ifGPblacktext
    \GPblacktexttrue
  }{}%
  % define a \g@addto@macro without @ in the name:
  \let\gplgaddtomacro\g@addto@macro
  % define empty templates for all commands taking text:
  \gdef\gplbacktext{}%
  \gdef\gplfronttext{}%
  \makeatother
  \ifGPblacktext
    % no textcolor at all
    \def\colorrgb#1{}%
    \def\colorgray#1{}%
  \else
    % gray or color?
    \ifGPcolor
      \def\colorrgb#1{\color[rgb]{#1}}%
      \def\colorgray#1{\color[gray]{#1}}%
      \expandafter\def\csname LTw\endcsname{\color{white}}%
      \expandafter\def\csname LTb\endcsname{\color{black}}%
      \expandafter\def\csname LTa\endcsname{\color{black}}%
      \expandafter\def\csname LT0\endcsname{\color[rgb]{1,0,0}}%
      \expandafter\def\csname LT1\endcsname{\color[rgb]{0,1,0}}%
      \expandafter\def\csname LT2\endcsname{\color[rgb]{0,0,1}}%
      \expandafter\def\csname LT3\endcsname{\color[rgb]{1,0,1}}%
      \expandafter\def\csname LT4\endcsname{\color[rgb]{0,1,1}}%
      \expandafter\def\csname LT5\endcsname{\color[rgb]{1,1,0}}%
      \expandafter\def\csname LT6\endcsname{\color[rgb]{0,0,0}}%
      \expandafter\def\csname LT7\endcsname{\color[rgb]{1,0.3,0}}%
      \expandafter\def\csname LT8\endcsname{\color[rgb]{0.5,0.5,0.5}}%
    \else
      % gray
      \def\colorrgb#1{\color{black}}%
      \def\colorgray#1{\color[gray]{#1}}%
      \expandafter\def\csname LTw\endcsname{\color{white}}%
      \expandafter\def\csname LTb\endcsname{\color{black}}%
      \expandafter\def\csname LTa\endcsname{\color{black}}%
      \expandafter\def\csname LT0\endcsname{\color{black}}%
      \expandafter\def\csname LT1\endcsname{\color{black}}%
      \expandafter\def\csname LT2\endcsname{\color{black}}%
      \expandafter\def\csname LT3\endcsname{\color{black}}%
      \expandafter\def\csname LT4\endcsname{\color{black}}%
      \expandafter\def\csname LT5\endcsname{\color{black}}%
      \expandafter\def\csname LT6\endcsname{\color{black}}%
      \expandafter\def\csname LT7\endcsname{\color{black}}%
      \expandafter\def\csname LT8\endcsname{\color{black}}%
    \fi
  \fi
  \setlength{\unitlength}{0.0500bp}%
  \begin{picture}(9936.00,2770.00)%
    \gplgaddtomacro\gplbacktext{%
      \csname LTb\endcsname%
      \put(215,28){\makebox(0,0)[r]{\strut{}-1}}%
      \put(215,657){\makebox(0,0)[r]{\strut{}-0.5}}%
      \put(215,1287){\makebox(0,0)[r]{\strut{} 0}}%
      \put(215,1916){\makebox(0,0)[r]{\strut{} 0.5}}%
      \put(215,2545){\makebox(0,0)[r]{\strut{} 1}}%
      \put(347,-192){\makebox(0,0){\strut{}}}%
      \put(1123,-192){\makebox(0,0){\strut{}}}%
      \put(1900,-192){\makebox(0,0){\strut{}}}%
      \put(2676,-192){\makebox(0,0){\strut{}}}%
      \put(3452,-192){\makebox(0,0){\strut{}}}%
    }%
    \gplgaddtomacro\gplfronttext{%
    }%
    \gplgaddtomacro\gplbacktext{%
      \csname LTb\endcsname%
      \put(215,28){\makebox(0,0)[r]{\strut{}-1}}%
      \put(215,657){\makebox(0,0)[r]{\strut{}-0.5}}%
      \put(215,1287){\makebox(0,0)[r]{\strut{} 0}}%
      \put(215,1916){\makebox(0,0)[r]{\strut{} 0.5}}%
      \put(215,2545){\makebox(0,0)[r]{\strut{} 1}}%
      \put(347,-192){\makebox(0,0){\strut{}}}%
      \put(1123,-192){\makebox(0,0){\strut{}}}%
      \put(1900,-192){\makebox(0,0){\strut{}}}%
      \put(2676,-192){\makebox(0,0){\strut{}}}%
      \put(3452,-192){\makebox(0,0){\strut{}}}%
    }%
    \gplgaddtomacro\gplfronttext{%
    }%
    \gplgaddtomacro\gplbacktext{%
      \csname LTb\endcsname%
      \put(215,28){\makebox(0,0)[r]{\strut{}-1}}%
      \put(215,657){\makebox(0,0)[r]{\strut{}-0.5}}%
      \put(215,1287){\makebox(0,0)[r]{\strut{} 0}}%
      \put(215,1916){\makebox(0,0)[r]{\strut{} 0.5}}%
      \put(215,2545){\makebox(0,0)[r]{\strut{} 1}}%
      \put(347,-192){\makebox(0,0){\strut{}}}%
      \put(1123,-192){\makebox(0,0){\strut{}}}%
      \put(1900,-192){\makebox(0,0){\strut{}}}%
      \put(2676,-192){\makebox(0,0){\strut{}}}%
      \put(3452,-192){\makebox(0,0){\strut{}}}%
    }%
    \gplgaddtomacro\gplfronttext{%
    }%
    \gplgaddtomacro\gplbacktext{%
      \csname LTb\endcsname%
      \put(3395,28){\makebox(0,0)[r]{\strut{}}}%
      \put(3395,657){\makebox(0,0)[r]{\strut{}}}%
      \put(3395,1287){\makebox(0,0)[r]{\strut{}}}%
      \put(3395,1916){\makebox(0,0)[r]{\strut{}}}%
      \put(3395,2545){\makebox(0,0)[r]{\strut{}}}%
      \put(3527,-192){\makebox(0,0){\strut{}}}%
      \put(4303,-192){\makebox(0,0){\strut{}}}%
      \put(5079,-192){\makebox(0,0){\strut{}}}%
      \put(5855,-192){\makebox(0,0){\strut{}}}%
      \put(6631,-192){\makebox(0,0){\strut{}}}%
      \put(5116,2655){\makebox(0,0){\strut{}extended critical filter}}%
    }%
    \gplgaddtomacro\gplfronttext{%
    }%
    \gplgaddtomacro\gplbacktext{%
      \csname LTb\endcsname%
      \put(3395,28){\makebox(0,0)[r]{\strut{}}}%
      \put(3395,657){\makebox(0,0)[r]{\strut{}}}%
      \put(3395,1287){\makebox(0,0)[r]{\strut{}}}%
      \put(3395,1916){\makebox(0,0)[r]{\strut{}}}%
      \put(3395,2545){\makebox(0,0)[r]{\strut{}}}%
      \put(3527,-192){\makebox(0,0){\strut{}}}%
      \put(4303,-192){\makebox(0,0){\strut{}}}%
      \put(5079,-192){\makebox(0,0){\strut{}}}%
      \put(5855,-192){\makebox(0,0){\strut{}}}%
      \put(6631,-192){\makebox(0,0){\strut{}}}%
      \put(5116,2655){\makebox(0,0){\strut{}extended critical filter}}%
    }%
    \gplgaddtomacro\gplfronttext{%
    }%
    \gplgaddtomacro\gplbacktext{%
      \csname LTb\endcsname%
      \put(3395,28){\makebox(0,0)[r]{\strut{}}}%
      \put(3395,657){\makebox(0,0)[r]{\strut{}}}%
      \put(3395,1287){\makebox(0,0)[r]{\strut{}}}%
      \put(3395,1916){\makebox(0,0)[r]{\strut{}}}%
      \put(3395,2545){\makebox(0,0)[r]{\strut{}}}%
      \put(3527,-192){\makebox(0,0){\strut{}}}%
      \put(4303,-192){\makebox(0,0){\strut{}}}%
      \put(5079,-192){\makebox(0,0){\strut{}}}%
      \put(5855,-192){\makebox(0,0){\strut{}}}%
      \put(6631,-192){\makebox(0,0){\strut{}}}%
      \put(5116,2655){\makebox(0,0){\strut{}extended critical filter}}%
    }%
    \gplgaddtomacro\gplfronttext{%
    }%
    \gplgaddtomacro\gplbacktext{%
      \csname LTb\endcsname%
      \put(6574,28){\makebox(0,0)[r]{\strut{}}}%
      \put(6574,657){\makebox(0,0)[r]{\strut{}}}%
      \put(6574,1287){\makebox(0,0)[r]{\strut{}}}%
      \put(6574,1916){\makebox(0,0)[r]{\strut{}}}%
      \put(6574,2545){\makebox(0,0)[r]{\strut{}}}%
      \put(6706,-192){\makebox(0,0){\strut{}}}%
      \put(7482,-192){\makebox(0,0){\strut{}}}%
      \put(8259,-192){\makebox(0,0){\strut{}}}%
      \put(9035,-192){\makebox(0,0){\strut{}}}%
      \put(9811,-192){\makebox(0,0){\strut{}}}%
      \put(8296,2435){\makebox(0,0){\strut{}}}%
    }%
    \gplgaddtomacro\gplfronttext{%
    }%
    \gplgaddtomacro\gplbacktext{%
      \csname LTb\endcsname%
      \put(6574,28){\makebox(0,0)[r]{\strut{}}}%
      \put(6574,657){\makebox(0,0)[r]{\strut{}}}%
      \put(6574,1287){\makebox(0,0)[r]{\strut{}}}%
      \put(6574,1916){\makebox(0,0)[r]{\strut{}}}%
      \put(6574,2545){\makebox(0,0)[r]{\strut{}}}%
      \put(6706,-192){\makebox(0,0){\strut{}}}%
      \put(7482,-192){\makebox(0,0){\strut{}}}%
      \put(8259,-192){\makebox(0,0){\strut{}}}%
      \put(9035,-192){\makebox(0,0){\strut{}}}%
      \put(9811,-192){\makebox(0,0){\strut{}}}%
      \put(8296,2435){\makebox(0,0){\strut{}}}%
    }%
    \gplgaddtomacro\gplfronttext{%
    }%
    \gplgaddtomacro\gplbacktext{%
      \csname LTb\endcsname%
      \put(6574,28){\makebox(0,0)[r]{\strut{}}}%
      \put(6574,657){\makebox(0,0)[r]{\strut{}}}%
      \put(6574,1287){\makebox(0,0)[r]{\strut{}}}%
      \put(6574,1916){\makebox(0,0)[r]{\strut{}}}%
      \put(6574,2545){\makebox(0,0)[r]{\strut{}}}%
      \put(6706,-192){\makebox(0,0){\strut{}}}%
      \put(7482,-192){\makebox(0,0){\strut{}}}%
      \put(8259,-192){\makebox(0,0){\strut{}}}%
      \put(9035,-192){\makebox(0,0){\strut{}}}%
      \put(9811,-192){\makebox(0,0){\strut{}}}%
      \put(8296,2435){\makebox(0,0){\strut{}}}%
    }%
    \gplgaddtomacro\gplfronttext{%
    }%
    \gplbacktext
    \put(0,0){\includegraphics{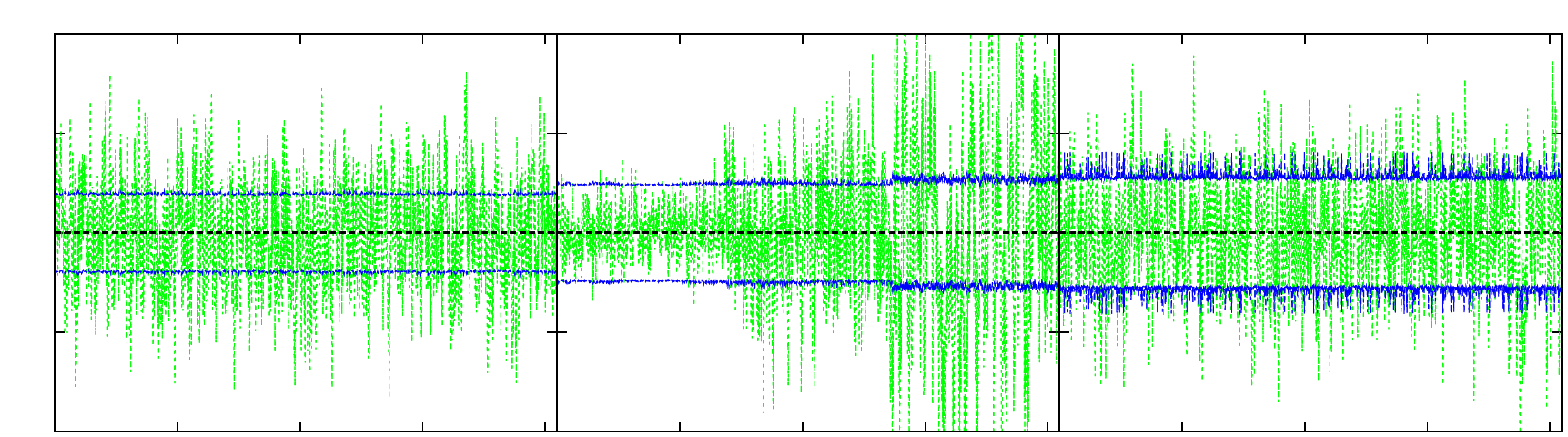}}%
    \gplfronttext
  \end{picture}%
\endgroup

%% file: noise-rec.bbl
\begin{thebibliography}{12}
\expandafter\ifx\csname natexlab\endcsname\relax\def\natexlab#1{#1}\fi

\bibitem[{{En{\ss}lin} \& {Frommert}(2011)}]{ensslin_frommert-2011}
{En{\ss}lin}, T.~A. \& {Frommert}, M., {Reconstruction of signals with unknown
  spectra in information field theory with parameter uncertainty},
  \href{http://dx.doi.org/10.1103/PhysRevD.83.105014}{\prd {~\bfseries 83}
  no.~10, (May 2011) 105014--+,},
  \href{http://arxiv.org/abs/1002.2928}{{\ttfamily arXiv:1002.2928
  [astro-ph.IM]}}

\bibitem[{{En{\ss}lin} {et~al.}(2009){En{\ss}lin}, {Frommert}, \&
  {Kitaura}}]{ensslin_frommert_kitaura-2009}
{En{\ss}lin}, T.~A., {Frommert}, M., \& {Kitaura}, F.~S., {Information field
  theory for cosmological perturbation reconstruction and nonlinear signal
  analysis}, \href{http://dx.doi.org/10.1103/PhysRevD.80.105005}{\prd
  {~\bfseries 80} no.~10, (Nov. 2009) 105005--+,},
  \href{http://arxiv.org/abs/0806.3474}{{\ttfamily arXiv:0806.3474}}

\bibitem[{{En{\ss}lin} \& {Weig}(2010)}]{ensslin_weig-2010}
{En{\ss}lin}, T.~A. \& {Weig}, C., {Inference with minimal Gibbs free energy in
  information field theory},
  \href{http://dx.doi.org/10.1103/PhysRevE.82.051112}{\pre {~\bfseries 82}
  no.~5, (Nov. 2010) 051112--+,},
  \href{http://arxiv.org/abs/1004.2868}{{\ttfamily arXiv:1004.2868
  [astro-ph.IM]}}

\bibitem[{{G{\'o}rski} {et~al.}(2005){G{\'o}rski}, {Hivon}, {Banday},
  {Wandelt}, {Hansen}, {Reinecke}, \& {Bartelmann}}]{gorski-2005}
{G{\'o}rski}, K.~M., {Hivon}, E., {Banday}, A.~J., {et~al.}, {HEALPix: A
  Framework for High-Resolution Discretization and Fast \ Analysis of Data
  Distributed on the Sphere}, \href{http://dx.doi.org/10.1086/427976}{\apj
  {~\bfseries 622} (Apr. 2005) 759--771,},
  \href{http://arxiv.org/abs/arXiv:astro-ph/0409513}{{\ttfamily
  arXiv:astro-ph/0409513}}

\bibitem[{Koltchinskii(2000)}]{koltchinskii-2000}
Koltchinskii, V.~I., {Empirical Geometry of Multivariate Data: A Deconvolution
  Approach}, The Annals of Statistics {~\bfseries 28} no.~2, (2000) 591--629,,
  \url{http://www.jstor.org/stable/2674043}

\bibitem[{Kullback \& Leibler(1951)}]{kullback-1951}
Kullback, S. \& Leibler, R.~A., On Information and Sufficiency, The Annals of
  Mathematical Statistics {~\bfseries 22} no.~1, (1951) pp. 79--86,,
  \url{http://www.jstor.org/stable/2236703}

\bibitem[{Olsen(1993)}]{olsen-1993}
Olsen, S.~I., Estimation of noise in images: an evaluation,
  \href{http://dx.doi.org/10.1006/cgip.1993.1022}{CVGIP: Graph. Models Image
  Process. {~\bfseries 55} (July 1993) 319--323,},
  \url{http://portal.acm.org/citation.cfm?id=167608.167619}

\bibitem[{{Oppermann} {et~al.}(2011){Oppermann}, {Junklewitz}, {Robbers}, \&
  {En{\ss}lin}}]{oppermann-2011a}
{Oppermann}, N., {Junklewitz}, H., {Robbers}, G., \& {En{\ss}lin}, T.~A.,
  {Probing magnetic helicity with synchrotron radiation and Faraday rotation},
  \href{http://dx.doi.org/10.1051/0004-6361/201015545}{\aap {~\bfseries 530}
  (Jun. 2011) A89+,}, \href{http://arxiv.org/abs/1008.1246}{{\ttfamily
  arXiv:1008.1246 [astro-ph.IM]}}

\bibitem[{{Oppermann}(2011)}]{oppermann-2011c}
{Oppermann}, N. e.~a., {An improved map of the Galactic Faraday depth (in
  prep.)}

\bibitem[{Sage(2010)}]{sage}
Sage. 2010, SAGE Mathematical Software, Version 4.3.3.,
  \url{http://www.sagemath.org}

\bibitem[{{Schwarz} \& {van Bellegem}(2010)}]{schwarz-2010}
{Schwarz}, M. \& {van Bellegem}, S., Consistent density deconvolution under
  partially known error distribution,
  \href{http://dx.doi.org/10.1016/j.spl.2009.10.012}{Statistics \& Probability
  Letters {~\bfseries 80} no.~3-4, (2010) 236 -- 241,},
  \url{http://www.sciencedirect.com/science/article/pii/S0167715209003988}

\bibitem[{{Starck} \& {Murtagh}(1998)}]{starck-1998}
{Starck}, J.-L. \& {Murtagh}, F., {Automatic Noise Estimation from the
  Multiresolution Support}, \href{http://dx.doi.org/10.1086/316124}{\pasp
  {~\bfseries 110} (Feb. 1998) 193--199,}

\end{thebibliography}
